%% file: main-janusdrop.tex
\documentclass[pre,showpacs,floats,twocolumn,floatfix,superscriptaddress,showkeys,letterpaper]{revtex4-1}
\setcitestyle{super}
\usepackage[T1]{fontenc}       
\usepackage{mathrsfs}
\usepackage{bm,amsbsy,amssymb,amsmath}
\usepackage{caption}
\usepackage{subcaption}
\usepackage{graphics,graphicx,dcolumn,fleqn,epic,eepic,float,tabularx}
\usepackage{multirow,rotate,rotating,color}
\usepackage[utf8]{inputenc}
\newcommand{\figref}[1]{Fig.~\ref{fig:#1}}
\newcommand{\eqnref}[1]{Eq.~(\ref{eq:#1})} 
  \definecolor{tuered}{RGB}{214,0,74}
  \definecolor{tueblue}{RGB}{0,102,204}
  
  \newcommand{\revisedtext}[1]{\textcolor{black}{#1}}

  \renewcommand{\vec}[1]{\mathbf{#1}}

\usepackage{tikz}
\usetikzlibrary{arrows}
\usetikzlibrary{arrows,decorations.markings}
\usepackage{relsize}
\tikzset{fontscale/.style = {font=\relsize{#1}}}
\usetikzlibrary{calc}

\begin{document}

\title{Direct Assembly of Magnetic Janus Particles at a Droplet Interface}
 \author{Qingguang Xie}
  \affiliation{Department of Applied Physics, Eindhoven University of Technology, P.O. Box 513, 5600MB Eindhoven, The Netherlands}
\author{Gary B. Davies}
\affiliation{St Paul's Girls' School, Brook Green, Hammersmith, London W6 7BS, United Kingdom} 
 \author{Jens Harting$^{*,\hspace{0.5mm}}$}
  \affiliation{Forschungszentrum J\"ulich, Helmholtz Institute Erlangen-N\"urnberg for Renewable Energy (IEK-11), F\"urther Stra{\ss}e 248, 90429 N\"urnberg, Germany.\\$^{*}$\hspace{0.8mm}j.harting@fz-juelich.de }
  \affiliation{Department of Applied Physics, Eindhoven University of Technology, P.O. Box 513, 5600MB Eindhoven, The Netherlands}
  \keywords{Janus particles, direct assembly, tunable deposition, colloids at interfaces, advanced printing techniques}
\begin{abstract}
\noindent
\revisedtext{\textbf{ABSTRACT:}} Self-assembly of nanoparticles at fluid-fluid interfaces is a promising route
to fabricate functional materials from the bottom-up. However, directing
and controlling particles into highly tunable and predictable structures --
while essential -- is a challenge. We present a liquid interface assisted
approach to fabricate nanoparticle structures with tunable properties. To
demonstrate its feasibility, we study magnetic Janus particles adsorbed at the
interface of a spherical droplet placed on a substrate. With an external magnetic field turned on, a
single particle moves to the location where its position vector relative to the
droplet centre is parallel to the direction of the applied field. Multiple
magnetic Janus particles arrange into reconfigurable hexagonal lattice
structures and can be directed to assemble at desirable locations on the
droplet interface by simply varying the magnetic field direction. We develop an
interface energy model to explain our observations, finding excellent agreement.
Finally, we demonstrate that the external magnetic field allows \revisedtext{one} to tune the
particle deposition pattern obtained when the droplet evaporates.  Our results
have implications for the fabrication of varied nanostructures on substrates
for use in nanodevices, organic electronics, or advanced display, printing and
coating applications.
\vspace{5mm}

\end{abstract}

\maketitle
\frenchspacing
\input{intro-janusdrop}

\input{single-particle}
\input{multi-particles2}

\input{final-janusdrop}
\input{methods}

\vspace{1mm}
\begin{acknowledgments}
  Financial support is acknowledged from NWO/STW (STW project 13291).
  We thank the J\"ulich Supercomputing Centre and the
  High Performance Computing Center Stuttgart for the technical support and allocated CPU time.
 \end{acknowledgments}
 \vspace{1mm}
\revisedtext{ 
\textit{\textbf{Supporting Information Available}}:
Movies visualizing the assembly process of multiple Janus particles are
available free of charge \textit{via} the Internet at http://pubs.acs.org.
}
\input{ref-acsnano.bbl}


\end{document}

%% file: intro-janusdrop.tex
The self-assembly of nanoparticles provides a promising
route to fabricate nano- and microstructured functional
materials.~\cite{Boal2000} In this type of self-assembly, the particles
organise into interesting structures due to particle-particle interactions such
as molecular bonds,~\cite{Olson2009} electromagnetic
interactions,~\cite{Hermanson2001} and capillary interactions. For
technological applications, we need the ability to direct and control the
particles so that they assemble into desirable structures, and
this ability remains one of the biggest challenges in
nanoscience.~\cite{Weiss2008,Grzelczak2010}

Fluid-fluid interfaces have been identified as a promising platform for the
controllable assembly of nanoparticles. Nanoparticles trap irreversibly at
fluid interfaces because they reduce the area of the energetically-costly
interface, and once absorbed, they try to arrange into their lowest energy
configurations.~\cite{Binks2001,Lin2005} For example, spherical CdSe
nanoparticles assemble into a disordered but densely packed monolayer to reduce
the total interfacial energy.~\cite{Lin2005} Moreover, going beyond the
canonical case of spherical particles with uniform surface properties to
particles that exhibit anisotropy (\revisedtext{\textit{e.g.,}} ellipsoids, rods) and non-uniform
surface properties (\revisedtext{\textit{e.g.,}} Janus-like particles) provides more varied behaviour.
For example, nanorods can be used to generate different packing structures
simply by varying their aspect ratio.~\cite{Boker2007}
However, until recently~\cite{Grzelczak2010,Jiang2011} it has been difficult to
direct and control the assembly process since the arrangements of particles
depended only on the intrinsic properties of the particles and interfaces and
not on properties that can be controlled on-the-fly.

The synthesis of anisotropic particles with specific physical properties (\revisedtext{\textit{e.g.,}}
electric or magnetic moments) that interact with an external magnetic or
electric field allows greater control of the self-assembly process at fluid
interfaces because we can easily vary the external field: the interfacial
energy of anisotropic particles at a fluid-fluid interface depends on their
orientation~\cite{Park2012,Gary2014a,Xie2015,Martin2016} and the external field
can be used to change the orientation of the particles. Recently, Davies \revisedtext{\textit{et al.}}~\cite{Gary2014b} investigated the behaviour of magnetic ellipsoidal
particles at flat fluid-fluid interfaces, and they demonstrated that the
capillary interactions between magnetic ellipsoidal particles can be tuned by
varying an external magnetic field, resulting in switchable, network-like
monolayer structures.  Similarly, Xie \revisedtext{\textit{et al.}}~\cite{Xie2016} showed that
spherical magnetic Janus particles adsorbed at flat fluid-fluid interfaces can
generate highly ordered, chain-like structures that can be controlled by
altering an external magnetic field.

Most research to date, however, has focussed on the behaviour of these
particles at flat fluid-fluid interfaces, a situation that is quite artificial
in the real world where most fluid-fluid interfaces exhibit some degree of
curvature. More recently, particles at curved fluid-fluid interfaces were used
to show interesting phenomena. Cavallaro \revisedtext{\textit{et al.}}~\cite{Cavallaro2011}
demonstrated that on a curved fluid interface, rod-like particles move to areas
of greatest curvature due to the interaction between particle induced
deformation and the natural curvature gradient of the interfaces.  Ershov \revisedtext{\textit{et al.}}~\cite{Ershov2013} showed that isotropic colloids organise into a square
pattern aligned along the principle curvature axis of a curved interface.
However, the interface curvature driven assembly approach investigated to date
relies on having quite complex control of the interface curvature, and is
therefore less relevant to the situation where particles assembly at an
isotropically-curved fluid interface, such as a spherical droplet interface.

In this paper, we study the behaviour of spherical magnetic Janus particles
adsorbed at a spherical droplet interface interacting with an external magnetic
field.  We show that we are able to direct the location of particles by varying
the magnetic field direction on this isotropically curved interface and exploit
the anisotropic surface properties of Janus particles. We demonstrate that the
particles assemble into a highly ordered hexagonal lattice-like monolayer at a
spherical droplet interface, which is drastically different from the structures
formed at a flat fluid-fluid interface.~\cite{Xie2016} 
Moreover, we demonstrate that by changing the direction of the magnetic field,
it is possible to tune the particle deposition pattern during the evaporation
of a Janus particle-laden droplet, thus providing a controllable platform for
the fabrication of desirable nanostructures on substrates for various
nanodevices.

%% file: single-particle.tex
\section*{Results and discussion}
\begin{figure}[htbp]
 \begin{center}
\begin{subfigure}{0.23\textwidth}
\vspace{0.5mm}
\includegraphics[width= 1.0\textwidth]{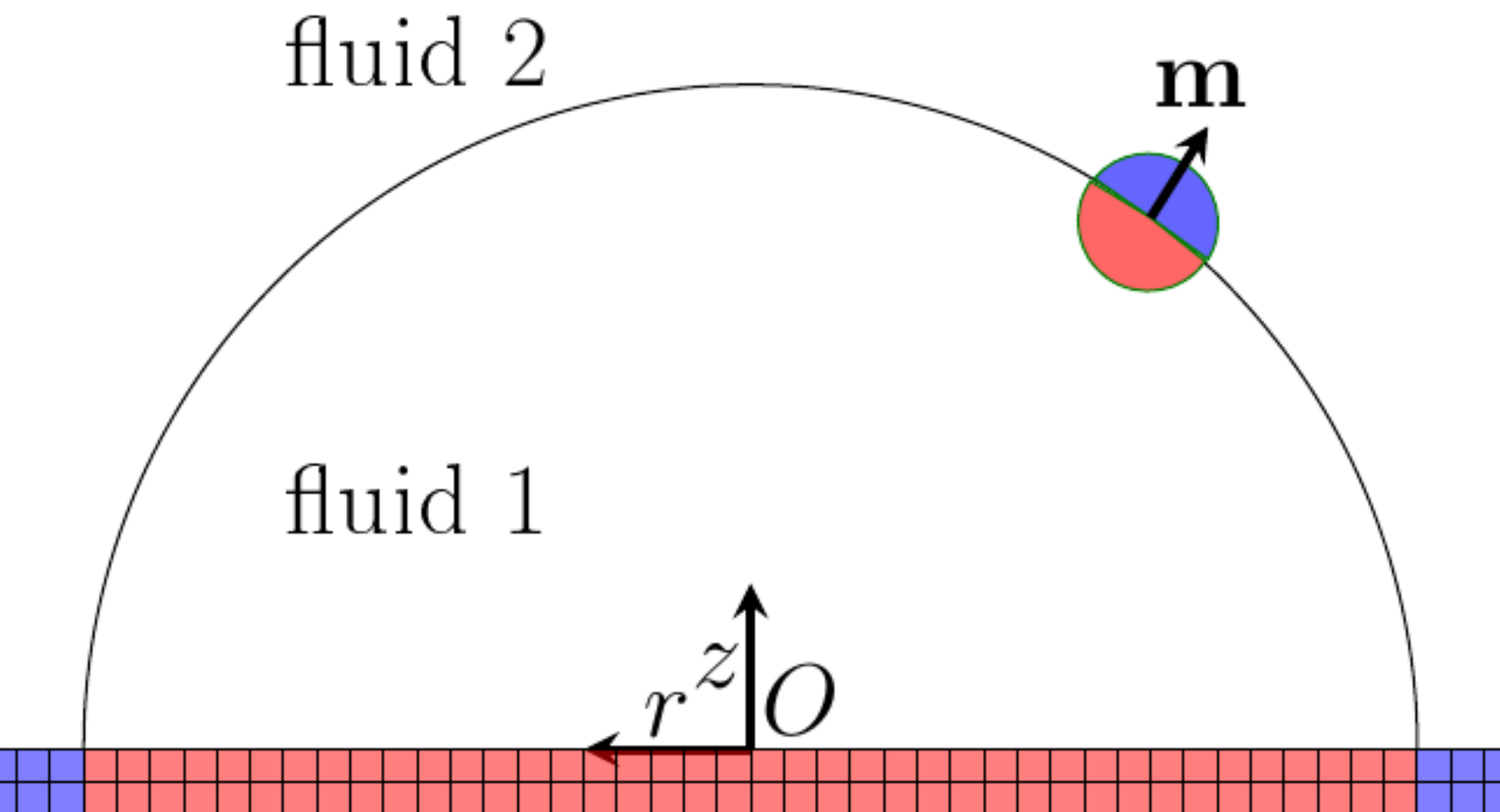}
\caption{}
\label{fig:equi}
\end{subfigure}
\begin{subfigure}{.23\textwidth}
 \includegraphics[width= 1.0\textwidth]{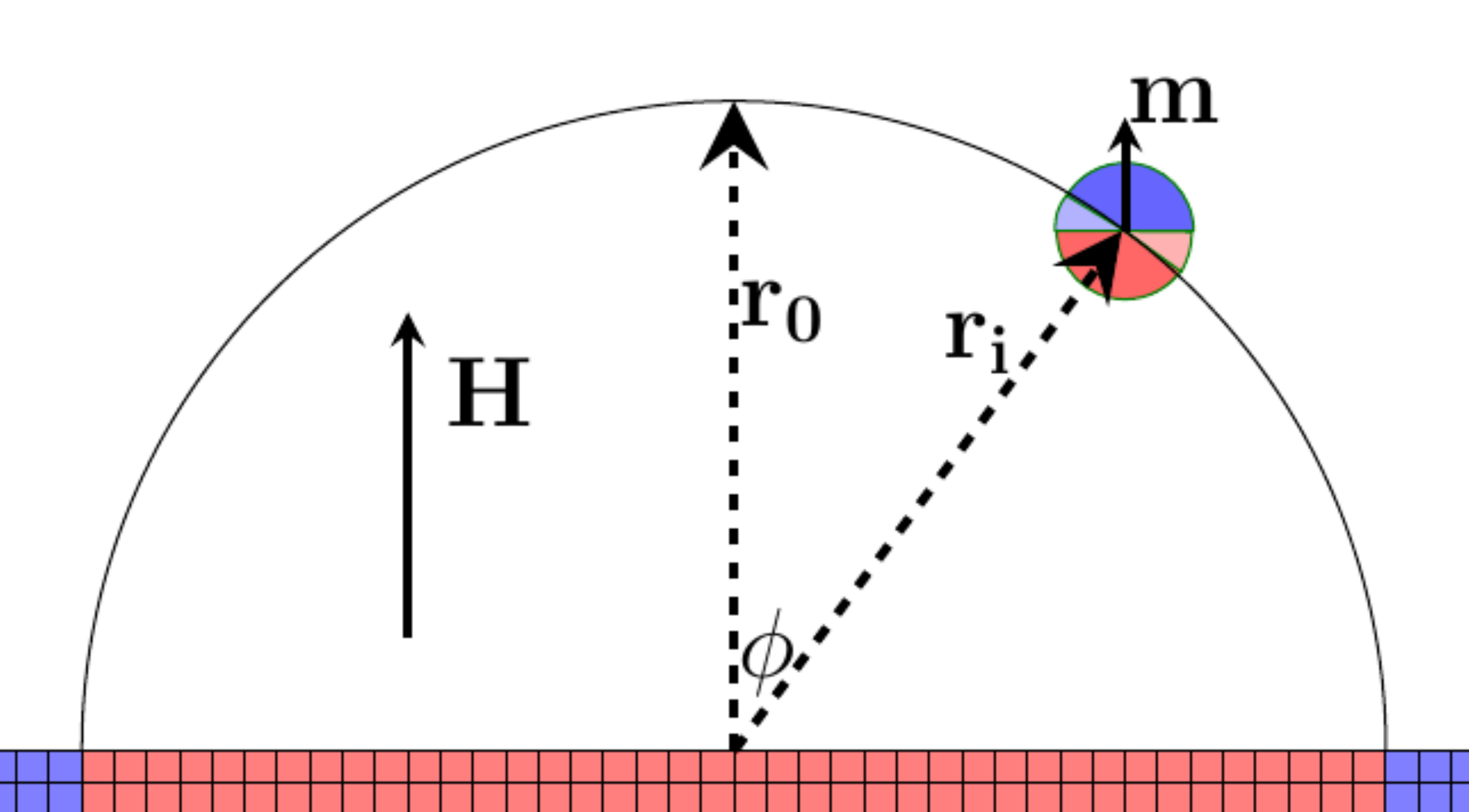}
\caption{}
\label{fig:tilt}
\end{subfigure}
 \end{center}
\caption{Side views of a single Janus particle adsorbed at a droplet interface
in its equilibrium orientation $(a)$ and under the influence of a vertical
magnetic field $(b)$. The Janus particle comprises an apolar and a polar
hemisphere. The particle's magnetic dipole moment $\mathbf{m}$ is orthogonal to
the Janus boundary, and the external magnetic field, $\mathbf{H}$, is directed
vertically upward.  The angle $\phi$ is defined as the angle between vector
$\mathbf{r}_0$ and particle position vector $\mathbf{r}_i$.}
\label{fig:janus-geo}
\end{figure}

\textbf{Geometry and surface properties.}
In order to understand the behaviour of multiple Janus particles adsorbed at a
surface droplet interface during evaporation, we first study the behaviour of a
single Janus particle. \figref{janus-geo} illustrates the system, which
comprises of a spherical magnetic Janus particle of radius $a$ adsorbed at the
interface of a droplet deposited on a chemically patterned substrate.  The
substrate covered by the droplet is hydrophilic with contact angle $30^{\circ}$
and the rest of the substrate is hydrophobic with contact angle $120^{\circ}$.
Thus, the droplet will be confined to the hydrophilic area with an effective
contact angle, and for simplicity we limit ourselves to investigating a droplet
of radius $R$ with an initial contact angle of $90^{\circ}$. We consider the
case of low Bond numbers in which the effect of gravity of the fluids is
negligible and the system is surface tension dominated, thus, the droplet has a
spherical shape with its centre located at $O$, as shown in~\figref{equi}.

The Janus particle comprises an apolar and a polar hemisphere with opposite
wettability, defined by the three-phase contact angles $\theta_A = 90^{\circ}
+\beta$  and $\theta_P = 90^{\circ} -\beta$, respectively, where $\beta$
represents the amphiphilicity of the particle. The boundary
between these two hemispheres is called the Janus boundary.  The particle's
magnetic dipole moment $\mathbf{m}$ is orthogonal to the Janus boundary, and
the applied external magnetic field is denoted by $\mathbf{H}$. We define a
dipole-field strength $B_m= |\mathbf{m}||\mathbf{H}|$, which represents the
magnitude of the interaction between the magnetic dipole and the external
magnetic field.  The angle between the particle's magnetic dipole vector and
the horizontal axis is denoted by $\alpha$.  We define a droplet field vector
$\mathbf{r}_0$ beginning at the droplet centre $O$ and ending at the droplet
interface that is directed parallel to the external magnetic field. We further
define a particle position vector $\mathbf{r}_i$ that points from the droplet
centre $O$ to the particle centre. Finally, the angle $\phi$ is defined as the
angle between droplet field vector $\mathbf{r}_0$ and particle position vector
$\mathbf{r}_i$.

In its equilibrium state, the two hemispheres of the Janus particle are fully
immersed in their preferred fluid, as shown in~\figref{equi}.  After switching
on an upward magnetic field, $\mathbf{H}$, the particle experiences a magnetic
torque that aligns it instantly with the dipole axis parallel to the field, as
shown in~\figref{tilt}.

\textbf{Single magnetic Janus particle.}

To simulate the behaviour of the Janus particle absorbed at the droplet
interface, we use a lattice Boltzmann method combined with a molecular dynamics
algorithm that has been developed and validated extensively in previous
research
projects,~\cite{Xie2015,Xie2016,DennisXieJens2016,Jansen2011,Frijters2012,KFGKH13}
which we briefly summarise in the Methods section. Initially, we place a
particle at the droplet interface with fixed orientation $\alpha=23.6^{\circ}$
vertically located at $z=0.4R$ and let the system equilibrate. After
equilibration, we release the particle and simultaneously switch on an upward
magnetic field. 

\begin{figure}[]
\centering
 \includegraphics[width= 0.45\textwidth]{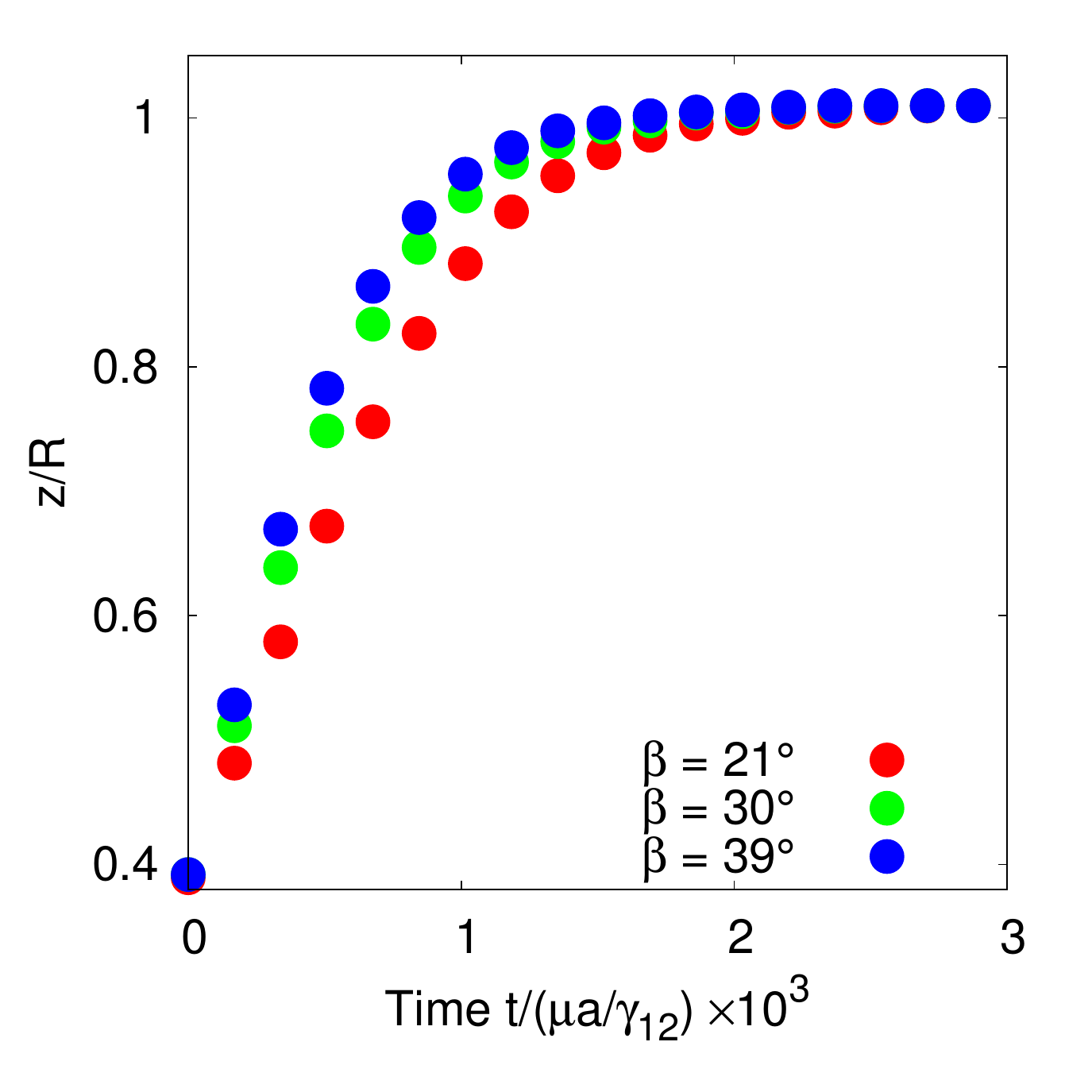}
 \caption{Time evolution of vertical position $z$ for particles with different amphiphilicities $\beta=21^{\circ}$ (red), $\beta=30^{\circ}$(cyan) and $\beta=39^{\circ}$(blue).}
\label{fig:p1-theta}
\end{figure}

\figref{p1-theta} shows time evolution of the vertical position $z$ for
particles with different amphiphilicities and dipole-field strengths $B_m/\pi
a^2 \gamma_{12}=1.31$.  The time is normalized by the viscosity of the fluid
$\mu$, particle radius $a$, and surface tension $\gamma_{12}$.  For all
amphiphilicities, the particle experiences a quick acceleration and moves
upward.  When the particle approaches the top of the droplet, it slows down and
finally remains at the droplet's top ($\phi=0$).  Moreover, we observe that
particles with larger amphiphilicities $\beta$ move faster just after the
magnetic field is applied.  This observation cannot be explained by theories
that suggest particles to move to areas of highest curvature~\cite{Ershov2013}
since the curvature of a spherical droplet interface is isotropic. 

To understand these observations, we develop a model describing the interface
energy of this system assuming that $(i)$ the interface deformation caused by
the nanometer-sized Janus particle is negligible,~\cite{Park2012} $(ii)$ and that the particle
radius is substantially smaller than the droplet radius.
To derive our model, we compare
the interface energy for the particle in its equilibrium state absorbed on the
droplet interface with its energy under the influence of an external magnetic
field. 

The interface energy for a particle in its equilibrium state (\figref{equi}) is 
\begin{equation}
 E_{\text{int}} = \gamma_{12}A_{12}^{ \text{int}} + \gamma_{a1}A_{a1}^{\text{int}} +\gamma_{p2}A_{p2}^{\text{int}}
  \mbox{.}
  \end{equation}
  Here, $\gamma_{ij}$ are the interface tensions between phases $i$ and
$j$ and $A_{ij}$ are the contact surface areas between phases $i$ and $j$,
where $i,j$ $=$ $\{1$: fluid, $2$: fluid, $a$: apolar, $p$: polar$\}$. 
The interface energy for a particle in an external magnetic field (\figref{tilt}) is 
\begin{eqnarray}
 E_{\text{mag}} = &&\gamma_{12}A_{12}^{\text{mag}} + \gamma_{a1}A_{a1}^{\text{mag}} + \gamma_{p2}A_{p2}^{\text{mag}} \nonumber \\ 
   +&&\gamma_{a2}A_{a2} + \gamma_{p1}A_{p1}
  \mbox{.}
\end{eqnarray}
The interface energy difference between the magnetic field induced orientation state and the initial equilibrium state is
\begin{equation}
\Delta E = 4 \phi a^2 \gamma_{12} \sin \beta
\mbox{.}
 \label{eq:edrop}
\end{equation}
The detailed introduction to this model is presented in the Methods section.

The interaction energy $\Delta E$ in \eqnref{edrop} is proportional to the
particle position angle $\phi$. Thus, minimising the interface energy requires
that the particle moves to the top of the droplet where $\phi=0$, in agreement
with our simulation results.  
Our theoretical model also predicts that particles with larger amphiphilicities cause a larger interface energy jump, 
explaining why the particle with higher amphiphilicity moves faster to the top of the droplet, as observed in our simulations.

We note that the magnetic Janus particles will relocate when we vary the direction of the magnetic field,
which has potential applications in sensor or display technology.
To estimate the responsive time of the magnetic Janus particles, 
we assume that a nanoparticle with radius $a=10$ $nm$ is adsorbed
at an oil-water droplet interface of radius $R=100$ $nm$, with surface tension $\gamma_{12}=70$ $\mathrm{mN/m}$,
and viscosity $\mu=1.0$ $mPa\cdot s$.~\cite{Duan2004,Lu2014}
Based on our simulation results shown in~\figref{p1-theta}, the normalized time $t/(\mu a / \gamma_{12})$ 
needed for the particle to relocate to the droplet top is $\approx 3\times 10^{3}$.
Therefore, the estimated response time is around $(\mu a / \gamma_{12}) \times 3 \times 10^3 \approx 4\times 10^{-7}s \approx 400ns$,
which is fast enough to satisfy the requirements of responsive materials
for advanced sensor or display technology.~\cite{Bai2014}
%
For a possible experimental relization of our system, we note that 
surface nanodroplets can be produced through a solvent exchange process in experiments.~\cite{Detlef2015,Zhang2017}

%% file: multi-particles2.tex

\begin{figure}[htbp]
 \begin{subfigure}{.2\textwidth}
\includegraphics[width= 1.0\textwidth]{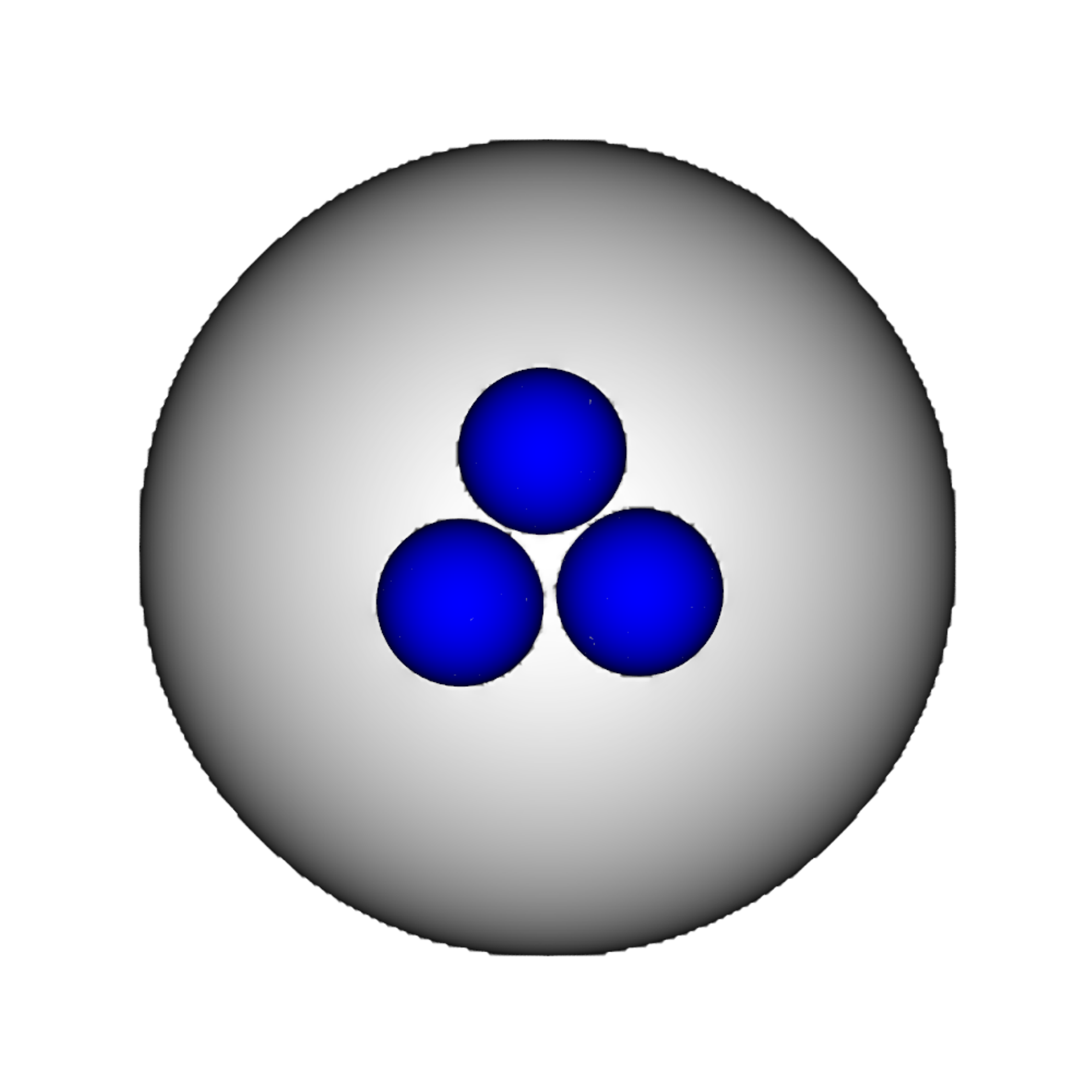}
 \subcaption{}
 \label{fig:p3}
\end{subfigure}
\hspace{1mm}
 \begin{subfigure}{.2\textwidth}
\includegraphics[width= 1.0\textwidth]{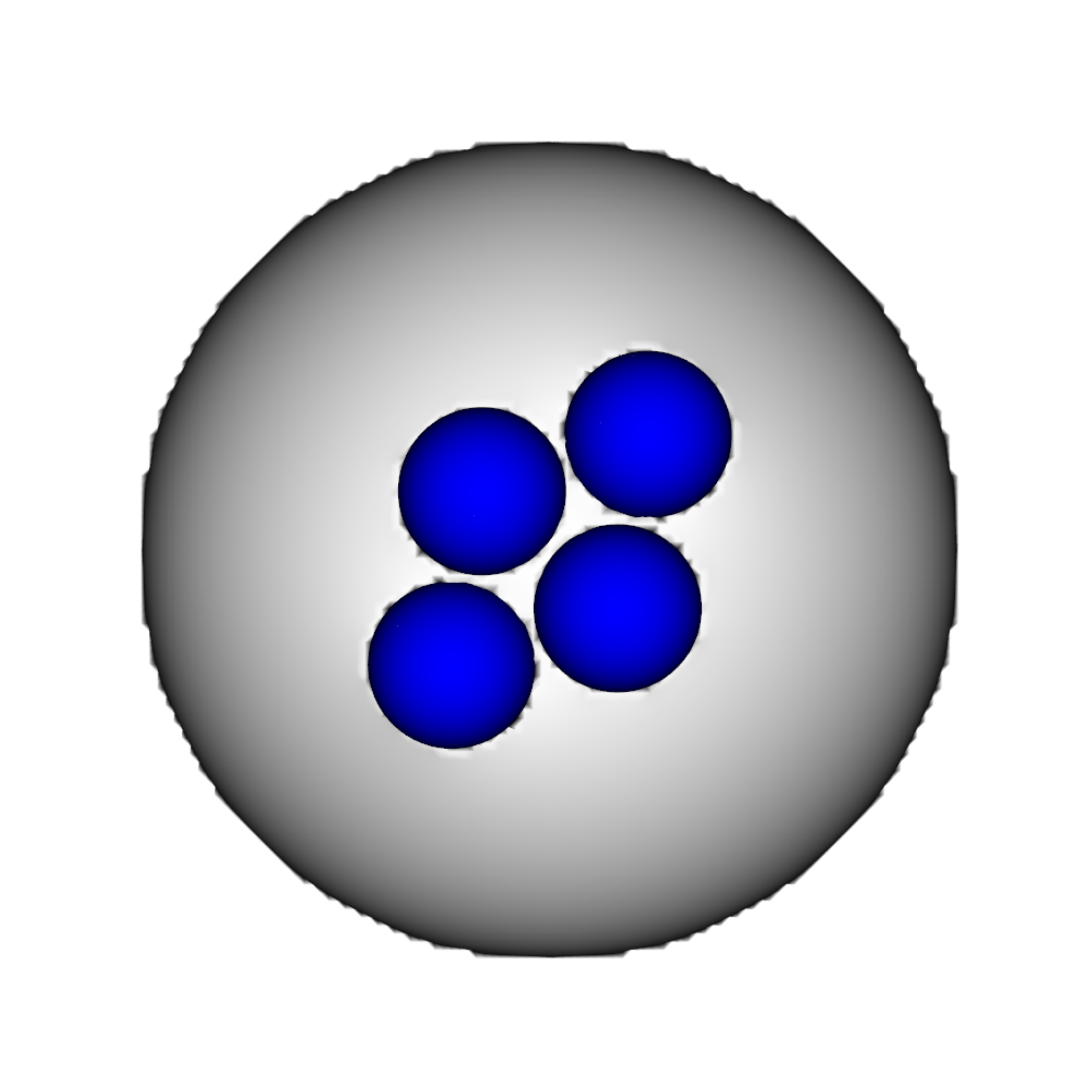}
 \subcaption{}
 \label{fig:p4}
\end{subfigure}
\\
\begin{subfigure}{.2\textwidth}
\includegraphics[width= 1.0\textwidth]{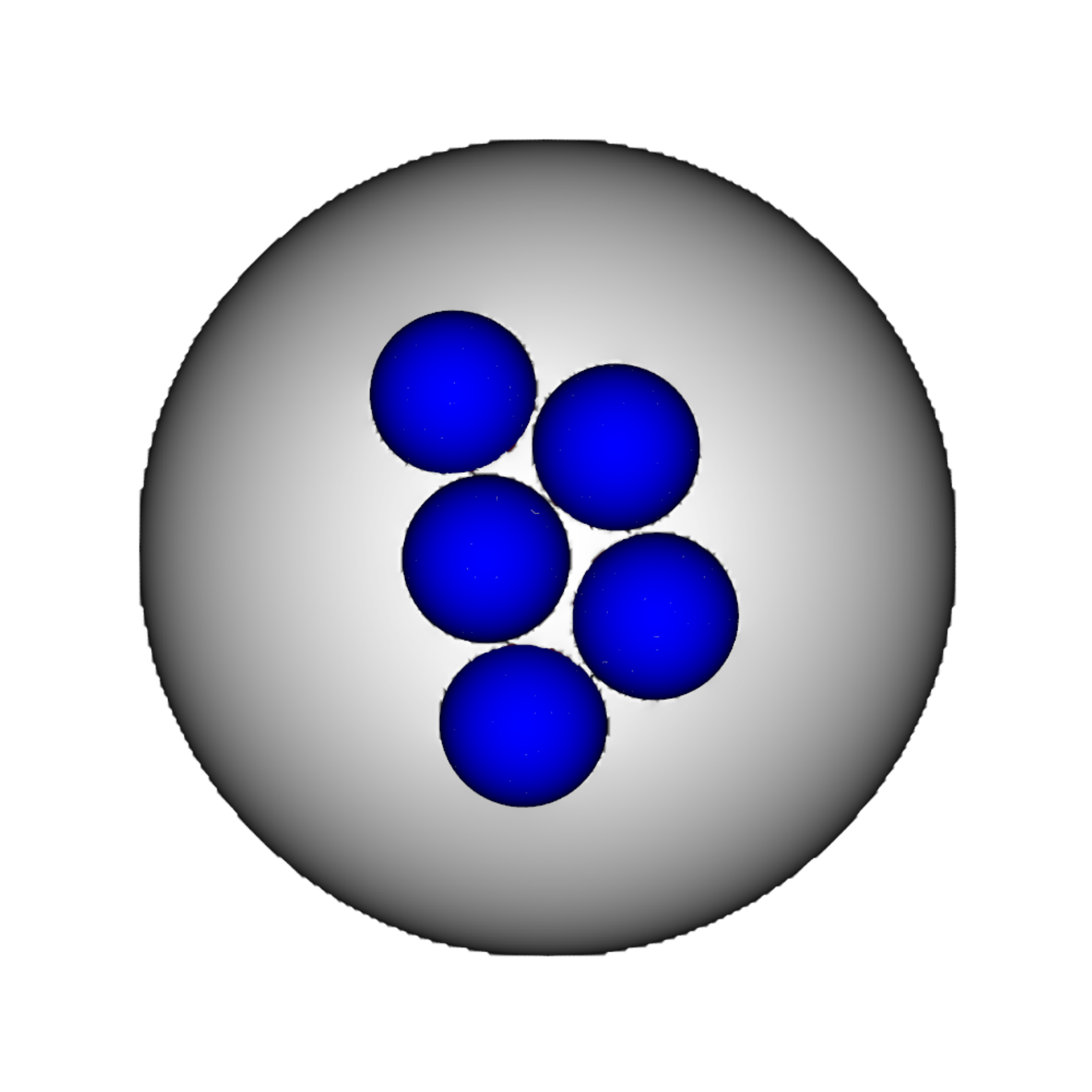}
 \subcaption{}
 \label{fig:p5}
 \end{subfigure}
 \hspace{1mm}
\begin{subfigure}{.2\textwidth}
\includegraphics[width= 1.0\textwidth]{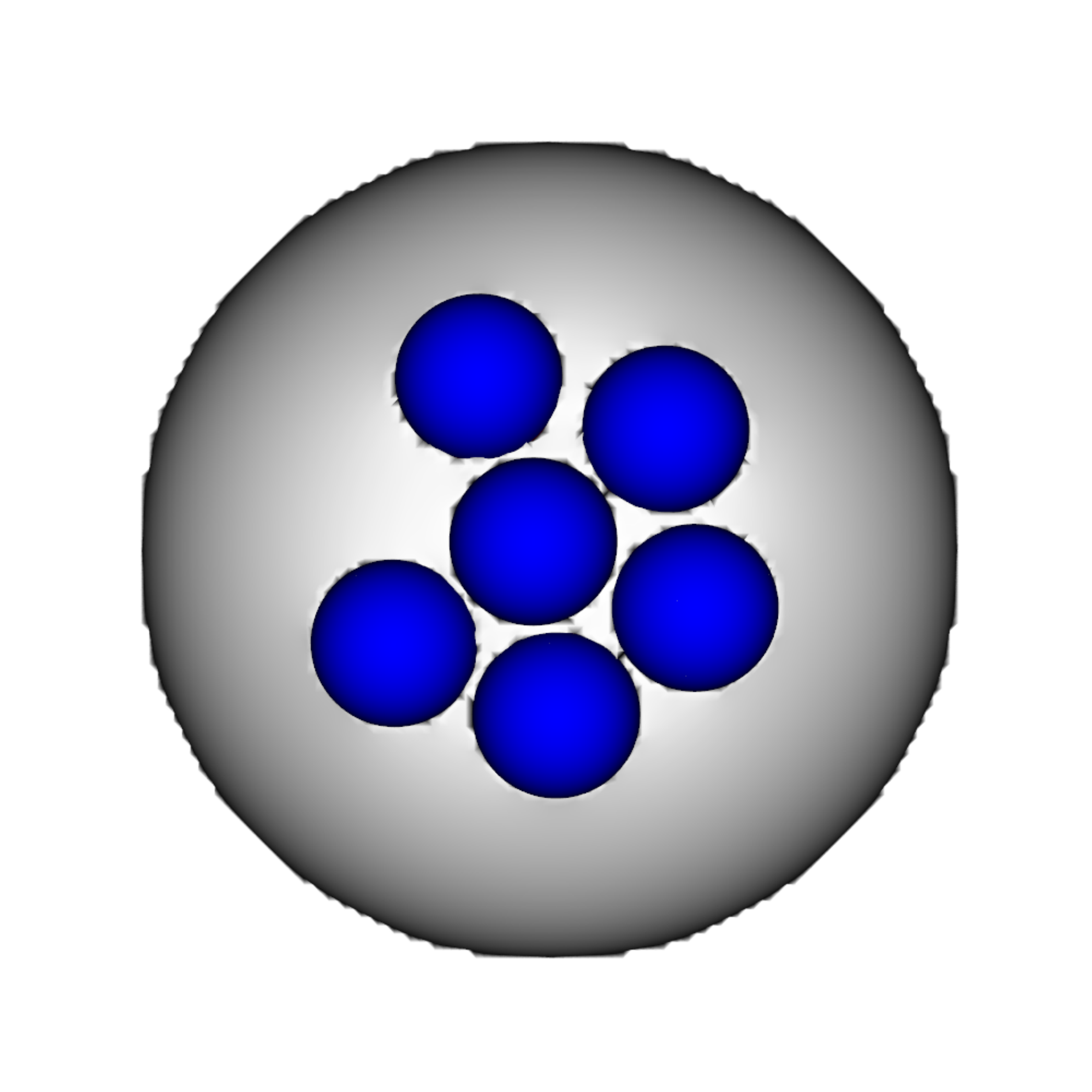}
 \subcaption{}
 \label{fig:p6}
 \end{subfigure}
 
\caption{Clusters observed in simulations for $N=3-6$ particles with
amphiphilicty $\beta=30^{\circ}$ and external dipole-field strength $B_m/\pi
a^2 \gamma_{12}=1.31$. The particles exhibit a hexagonal lattice-like
arrangement.}
\label{fig:p36-snap}
\end{figure}

\textbf{Multiple magnetic Janus particles.}
%
Having obtained a good understanding of the behaviour of a single magnetic Janus particle adsorbed at a surface droplet interface, we investigate the self-assembly of multiple magnetic Janus particles.
Starting from a random initial placement
of $3$ magnetic Janus particles adsorbed at the interface, we let
the system reach the steady-state. We then add one more particle at the droplet interface, let the system reach steady state again and repeat this process for up to $N=6$ particles. 
We show the structures obtained in \figref{p36-snap} for a given number of particles.

$N=3$ particles arrange into a triangular lattice-like structure around the
droplet centre (\figref{p3}), instead of the straight chain-like configuration
found by magnetic Janus particles at flat fluid-fluid
interfaces.~\cite{Xie2016}  This structure is similar to the $3$-particle
structure observed for colloidal particles adsorbed at a liquid crystal droplet
interface~\cite{Rahimi2015} and for colloidal particles interacting with short
range-interactions.~\cite{Pine2003,Arkus2009}  For $N=4-6$ particles, we find
the particles all favour hexagonal arrangements (\figref{p4}, \figref{p5} and
\figref{p6}), which suggests the hexagonal arrangement may be the interface
energy minimum configuration for $N > 3$ particles. 

Applying our free energy model for a single particle to the case for $N$
particles with assuming that the magnetic dipole-dipole interactions between
the particles are negligible, the interface energy is:
\begin{equation}
 \Delta E_t = 4  a^2 \gamma_{12} \sin \beta \sum_{i}^{N} \phi_i
 \mbox{.}
\end{equation}

\begin{figure}[htbp]
\includegraphics[width= 0.45\textwidth]{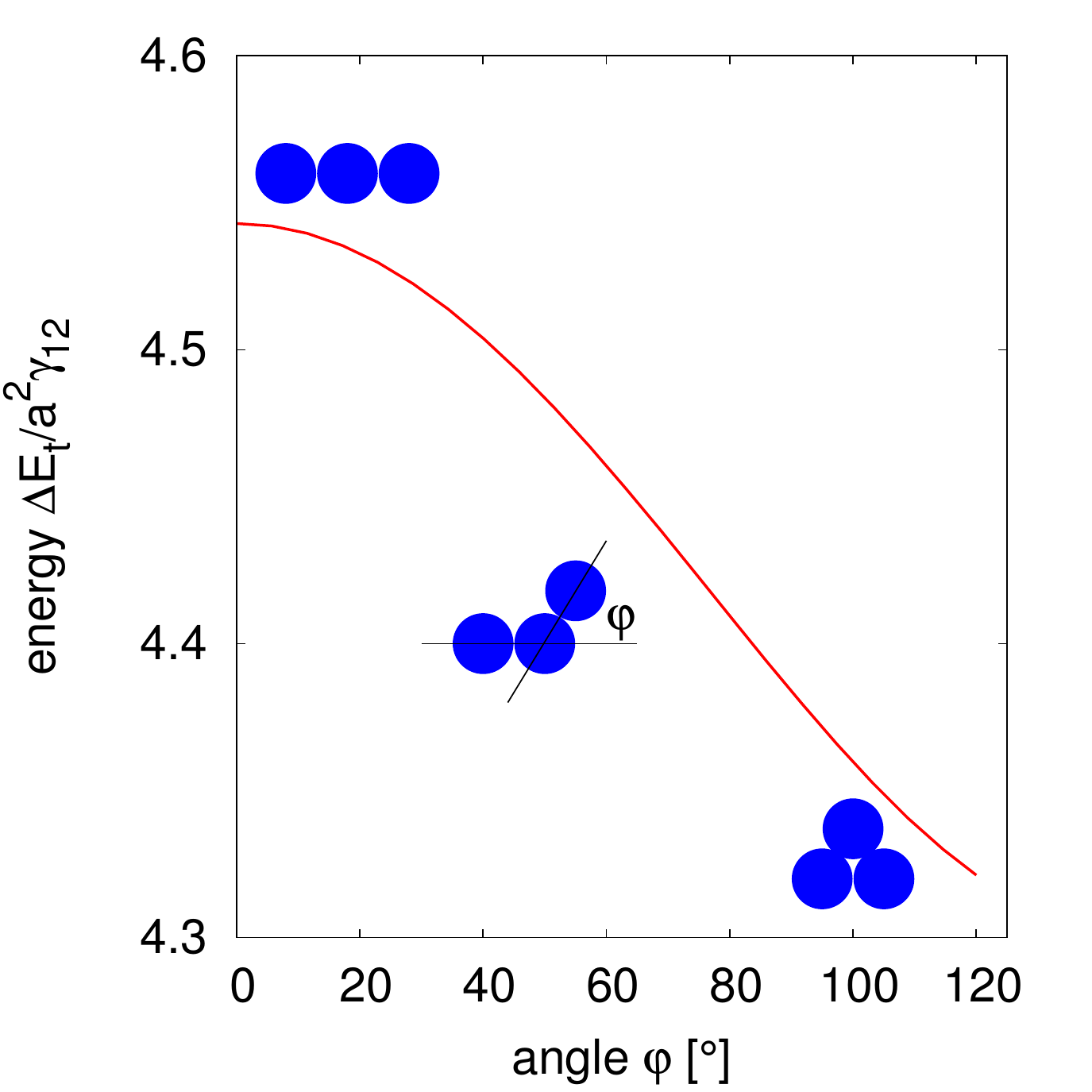}
\caption{Normalized interface energy $\Delta E_t/a^2 \gamma_{12} $ as a function
of particle pair-pair angle $\varphi$. The interface energy takes a minimum at
$\varphi = 120^{\circ}$, which represents a triangular lattice-like
arrangement, in agreement with our simulation results.}
\label{fig:p3-energy}
\end{figure}

To estimate the total interface energy for the $3$-particle system, we fix two
particles next to each other, and move the third particle with the pair angle
$\varphi$ from $\varphi=0^{\circ}$ to $\varphi=120^{\circ}$, as illustrated in
the inset of~\figref{p3-energy}.  \figref{p3-energy} shows that the interface
energy decreases monotonically with increasing particle pair-pair angle
$\varphi$ and the interface energy becomes minimal at $\varphi=120^{\circ}$,
which represents a triangular arrangement, in agreement with simulation
results.

For $N=4-6$ particles, it is nontrivial to estimate the interface energy as a
function of a single parameter, however, after some mathematical analysis we
find that minimising $\sum_{i}^{N} \phi_i$ is identical to minimising the total
particle-centre to droplet-centre distance $L = \sum_{i}^{N} |\mathbf{r}_i -
\mathbf{r}_0|$, with the following two constraints: $1)$ particles are located
at the droplet interface, $ |\mathbf{r}_i| =R $, and $2)$ particles do not
overlap, $|\mathbf{r}_i - \mathbf{r}_j| \geq 2a$.  In addition, in a strong
magnetic field, the orientation of the particle is fixed parallel to the
magnetic field direction and the particles are adsorbed at the droplet
interface, which leaves only two degrees of freedom for the particles.
Therefore, we can model our system as a quasi-$2D$ system.  Minimising $L$ for
spherical particles in a $2D$ system yields particles that conform to closest
packing, which is a hexagonal arrangement. Therefore, our theoretical model's
prediction that hexagonal arrangement is energetically favourable for
self-assembly of multiple magnetic Janus particles agrees with our simulation
results.

\begin{figure}[htbp]
\begin{subfigure}{.15\textwidth}
\includegraphics[width= 0.9\textwidth]{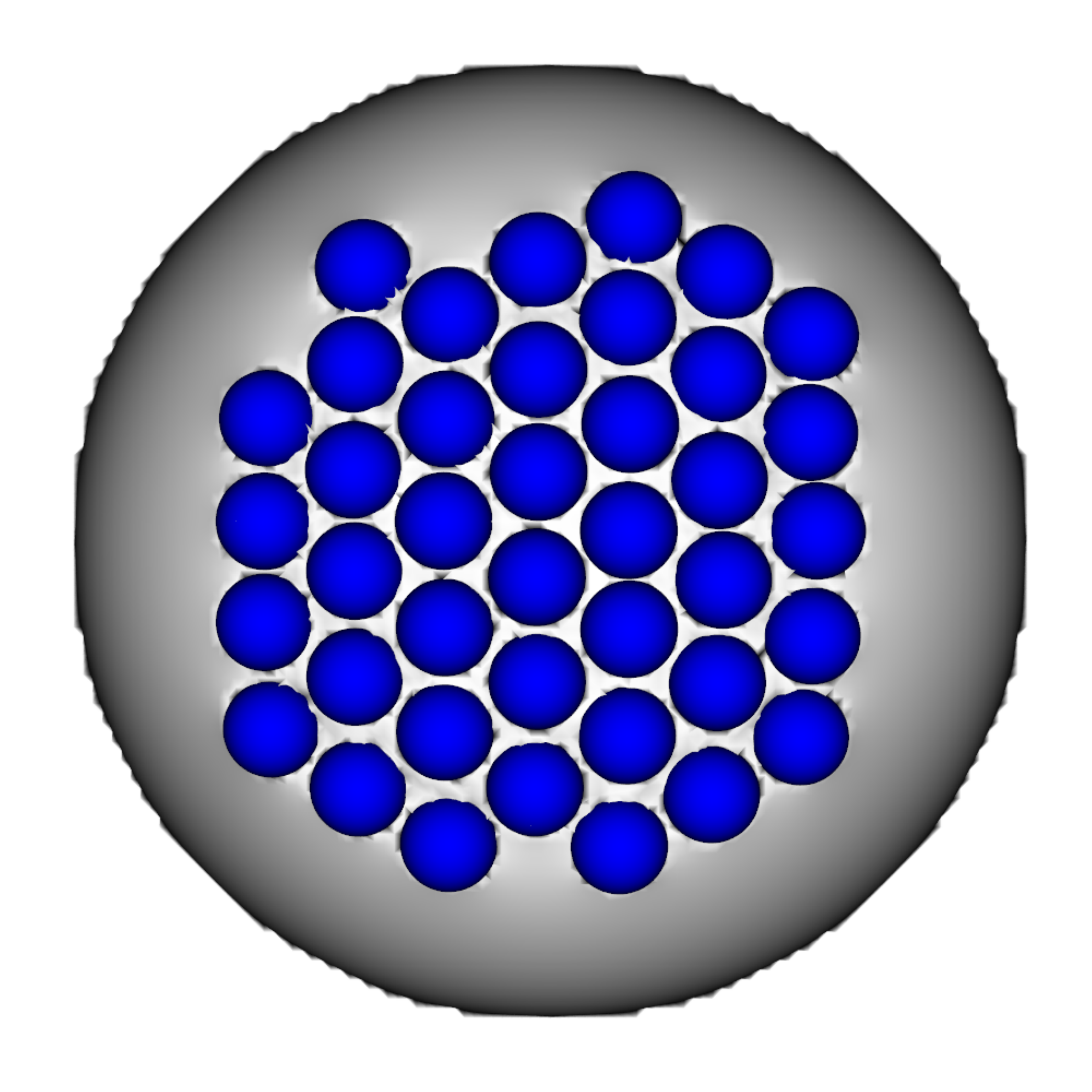}
 \label{fig:p40-up-top}
 \end{subfigure}
  \begin{subfigure}{.15\textwidth}
\includegraphics[width= 1.0\textwidth]{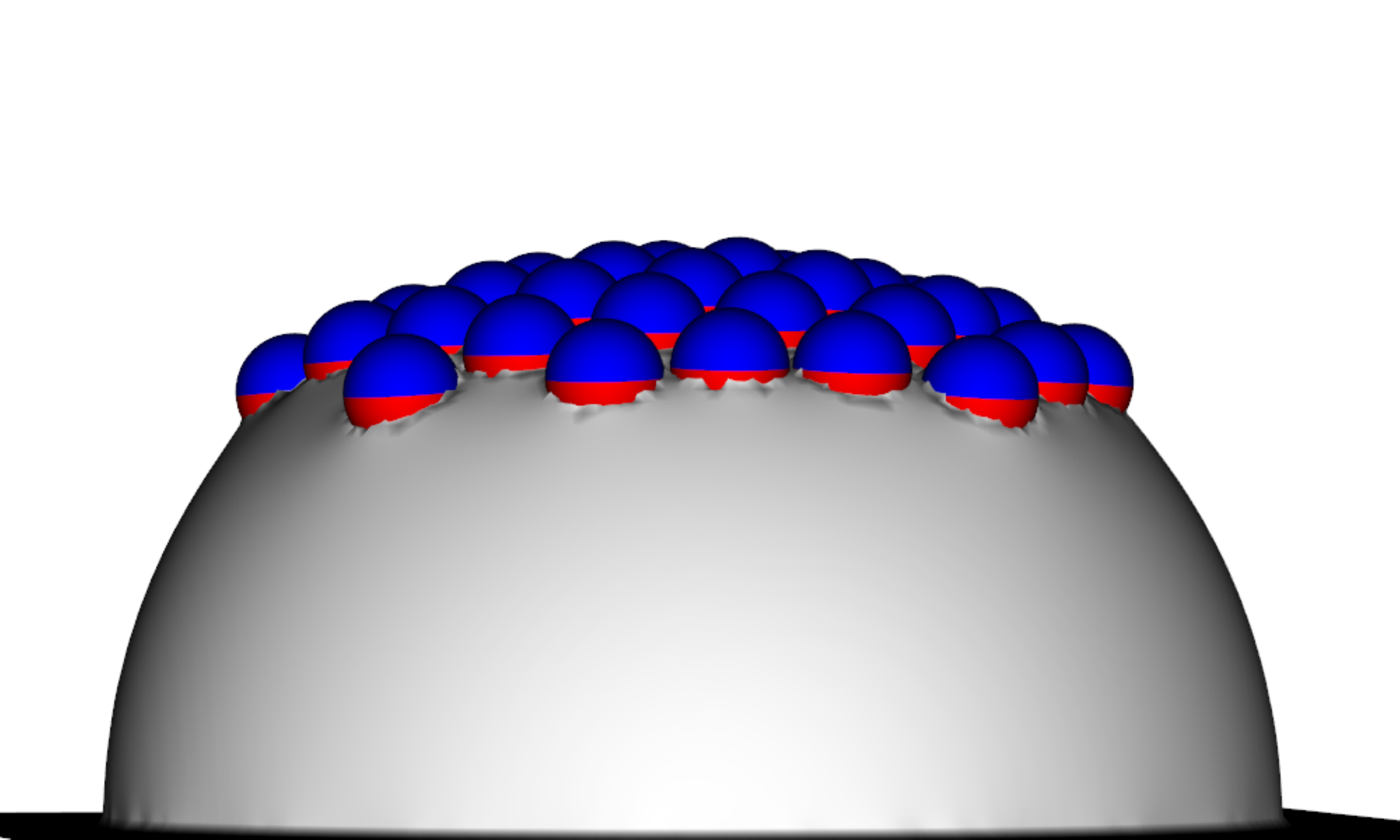}
 \label{fig:p40-up-side}
 \end{subfigure}
 \\
 \begin{subfigure}{.15\textwidth}
\includegraphics[width= 0.9\textwidth]{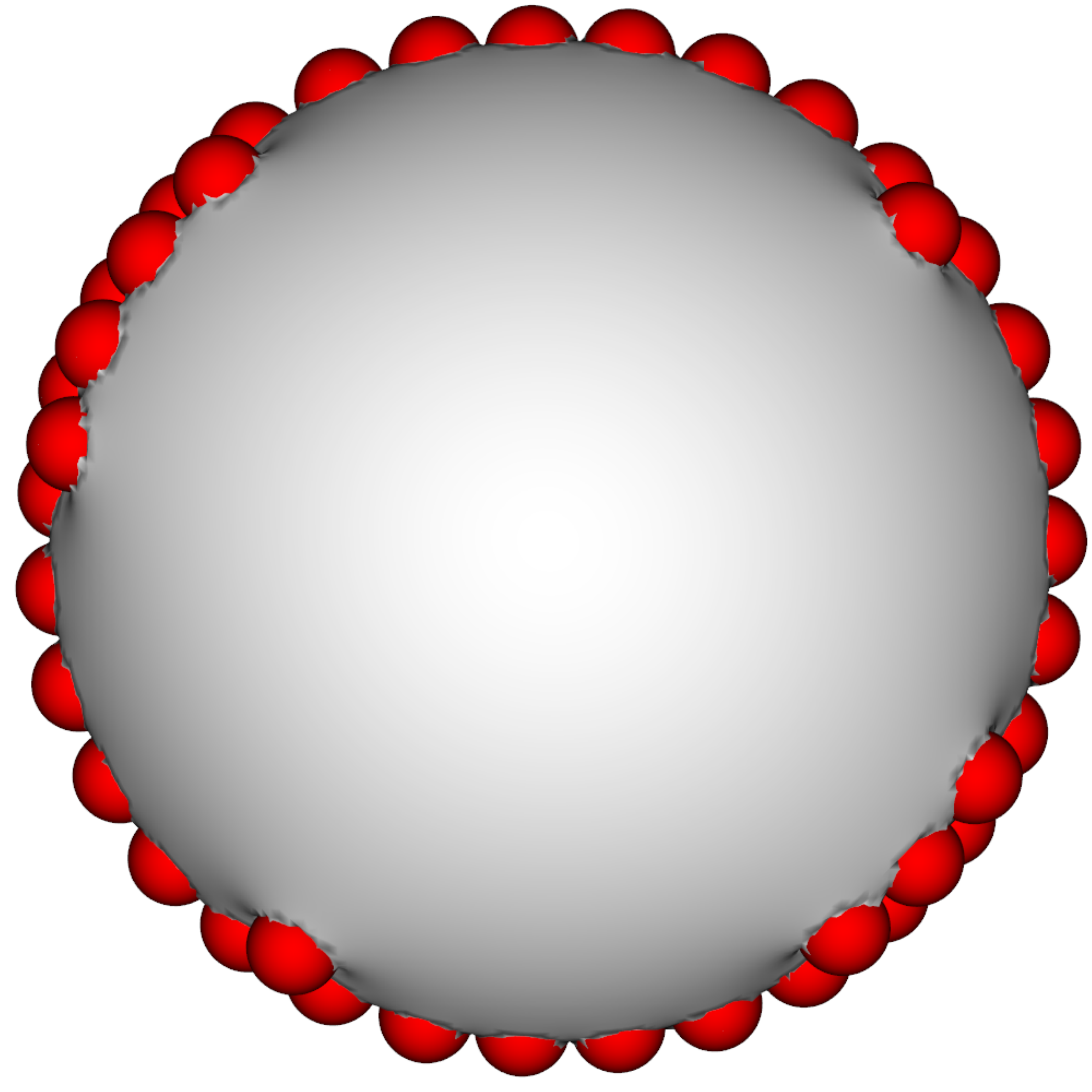}

 \label{fig:p40-down-top}
 \end{subfigure}
 \begin{subfigure}{.15\textwidth}
\includegraphics[width= 0.9\textwidth]{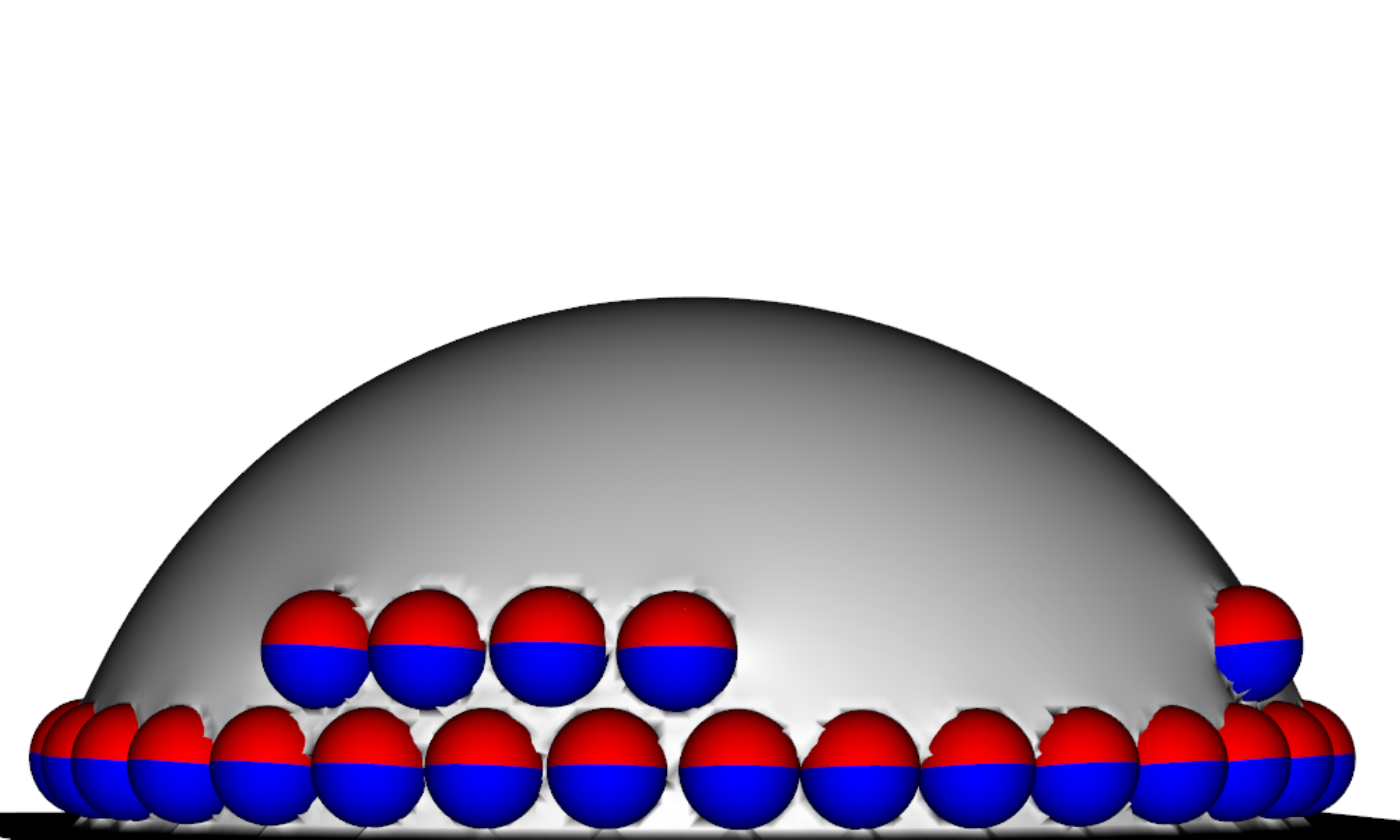}
 \label{fig:p40-down-side}
 \end{subfigure}
 \\
 \begin{subfigure}{.15\textwidth}
\includegraphics[width= 0.9\textwidth]{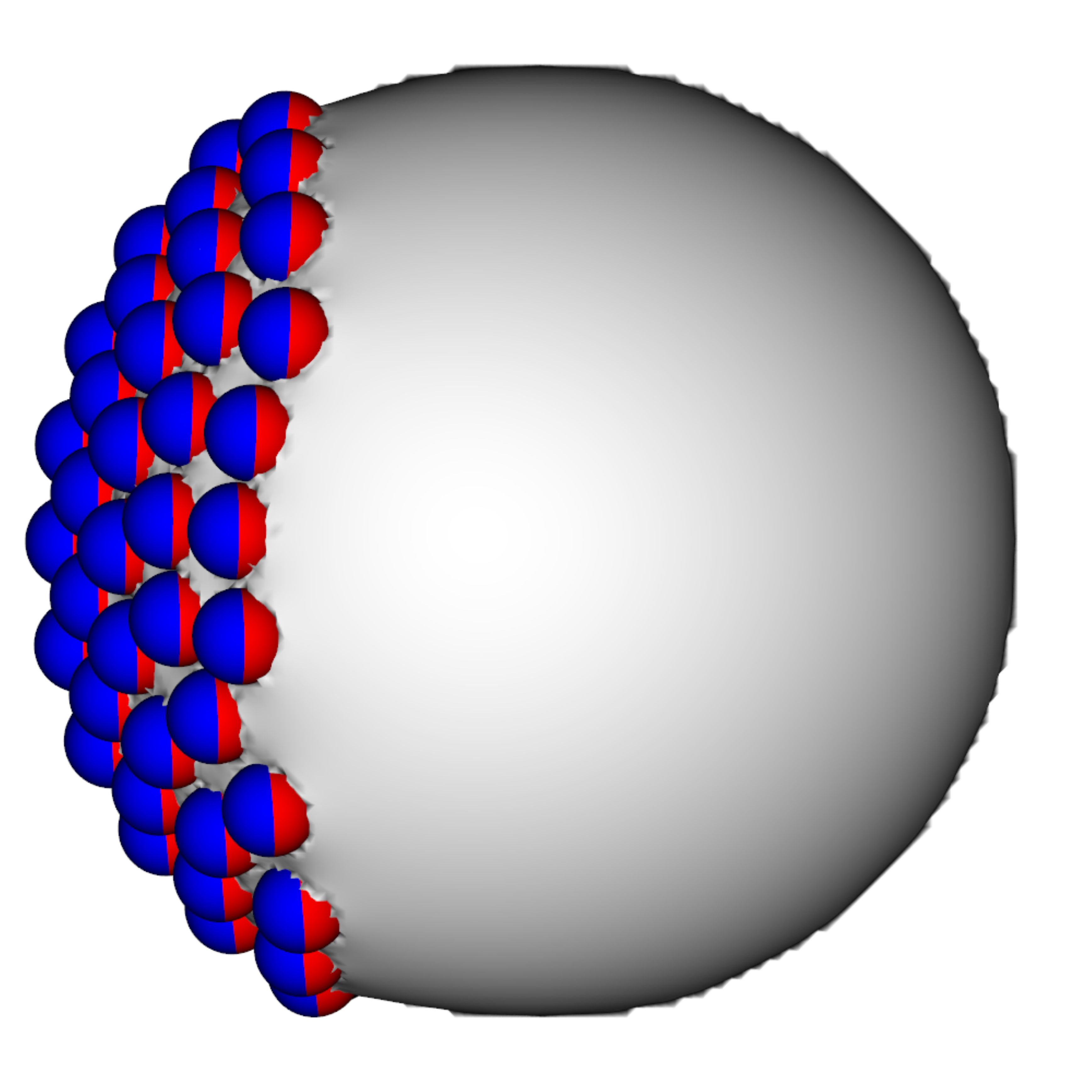}
 \label{fig:p40-left-top}
 \end{subfigure}
 \begin{subfigure}{.15\textwidth}
\includegraphics[width= 0.9\textwidth]{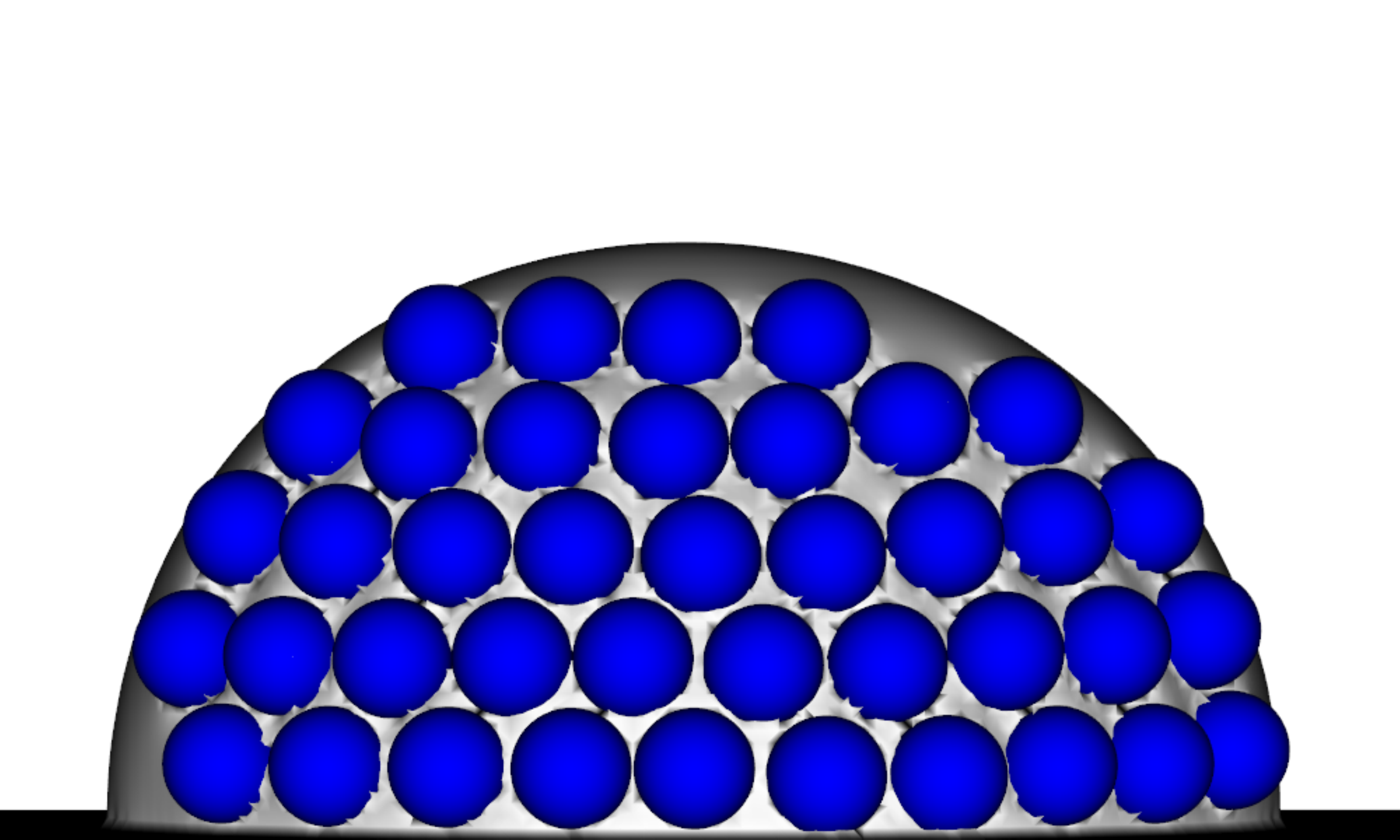}
 \label{fig:p40-left-side}
 \end{subfigure}
 
\caption{Snapshot of $N = 40$ particles adsorbed at a droplet interface with
amphiphilicty $\beta=30^{\circ}$ and dipole-field strength $B_m/\pi
a^2\gamma_{12}=1.31$ under different magnetic field directions. The particles
relocate and form different ordered structures depending on the direction of
the magnetic field.}
\label{fig:p40-snap}
\end{figure}

\textbf{Varying the magnetic field direction.}
With even more particles $N=40$, we investigate the relocation and
rearrangement of particles by varying the external field starting from pointing
vertically upward, to vertically downward and finally to left horizontal
direction. Under an upward magnetic field, the particles assemble at the top
of the droplet ordered in a hexagonal arrangement (\figref{p40-up-top} and
\figref{p40-up-side}). When we switch the magnetic field to the downward
direction, the particles rotate instantly by $180^{\circ}$, move immediately
downwards towards to the surface droplet contact line and finally align in a
ring-like structure at the edge of the droplet (\figref{p40-down-top} and
\figref{p40-down-side}). On switching the magnetic field to the left
horizontal direction, the particles rotate by $90^{\circ}$, move to the left side
of the droplet and form highly ordered layers (\figref{p40-left-top} and
\figref{p40-left-side}). [Movie S1]

Our simulations therefore show that it is possible to direct the particles to
assemble at desirable locations at the droplet interface based on the direction
of the external magnetic field, and that the final assembled structure is
tunable depending on the direction and magnitude of the applied magnetic field.
One can immediately forsee technological applications as \textit{e.g.,} a photonic
material or alternative to currently used electrophoretic/electronic inks (E-inks), where
electrophoretic particles encapsulated in a microdroplet are relocated by an
external electric field.~\cite{Comiskey1998}

\textbf{Particle deposition during evaporation.}
Finally, we investigate the case of many magnetic Janus particles adsorbed at a
surface droplet interface that undergoes evaporation, a scenario that occurs
frequently in reality when coatings and paints dry. To simulate fluid
evaporation, we use a diffusion dominated evaporation model, developed recently
in our group.~\cite{DennisXieJens2016} The starting configuration of the system
before evaporation is the same as those shown in \textit{e.g.,} Fig~\ref{fig:p40-snap}.
Once a steady-state is reached, we start to evaporate the droplet.

\begin{figure}[b!]
\begin{subfigure}{.10\textwidth}
\includegraphics[width= 1.0\textwidth]{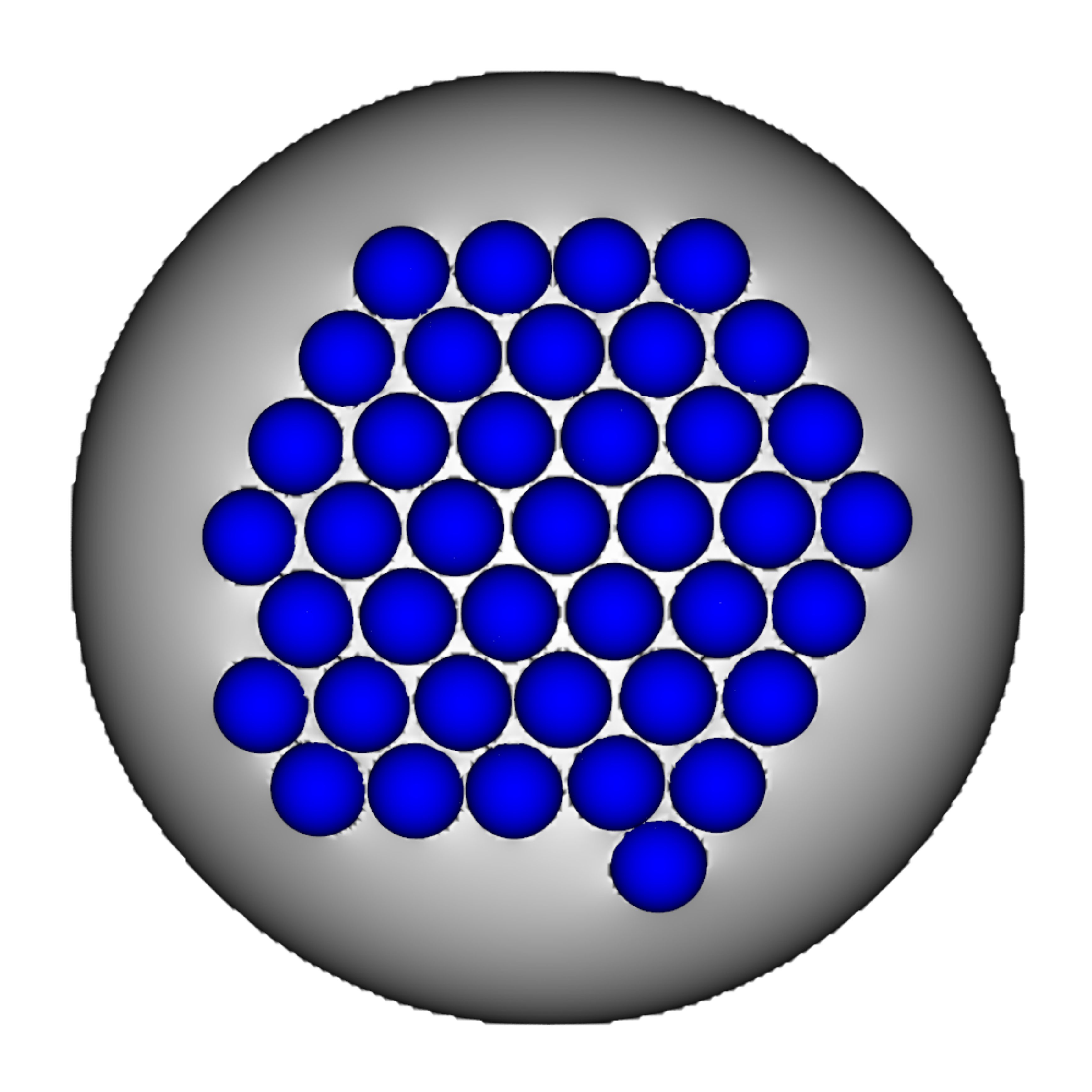}
 \label{fig:p30-zup-1}
 \end{subfigure}
 \begin{subfigure}{.10\textwidth}
\includegraphics[width= 1.0\textwidth]{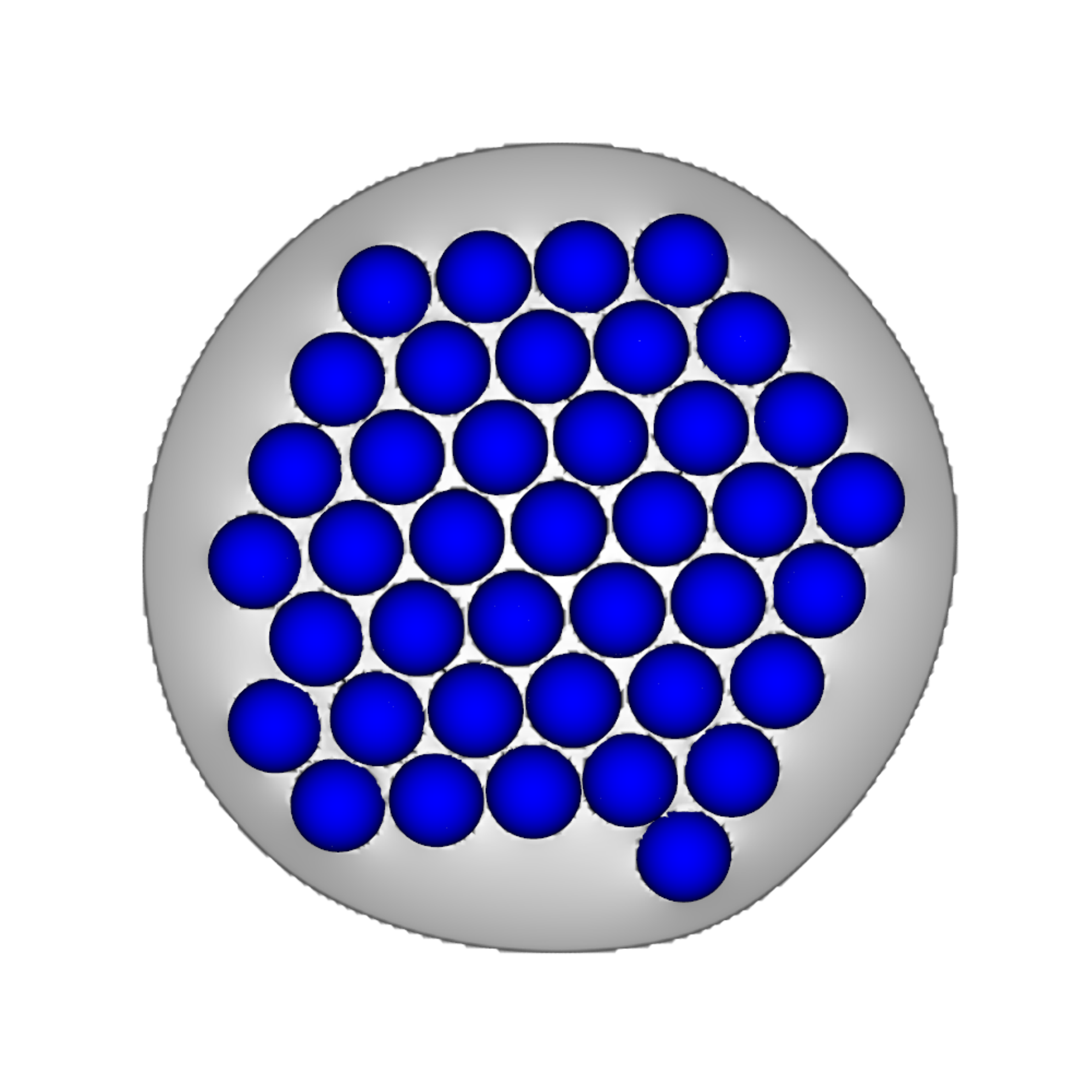}
 \label{fig:p30-zup-2}
\end{subfigure}
\begin{subfigure}{.1\textwidth}
\includegraphics[width= 1.0\textwidth]{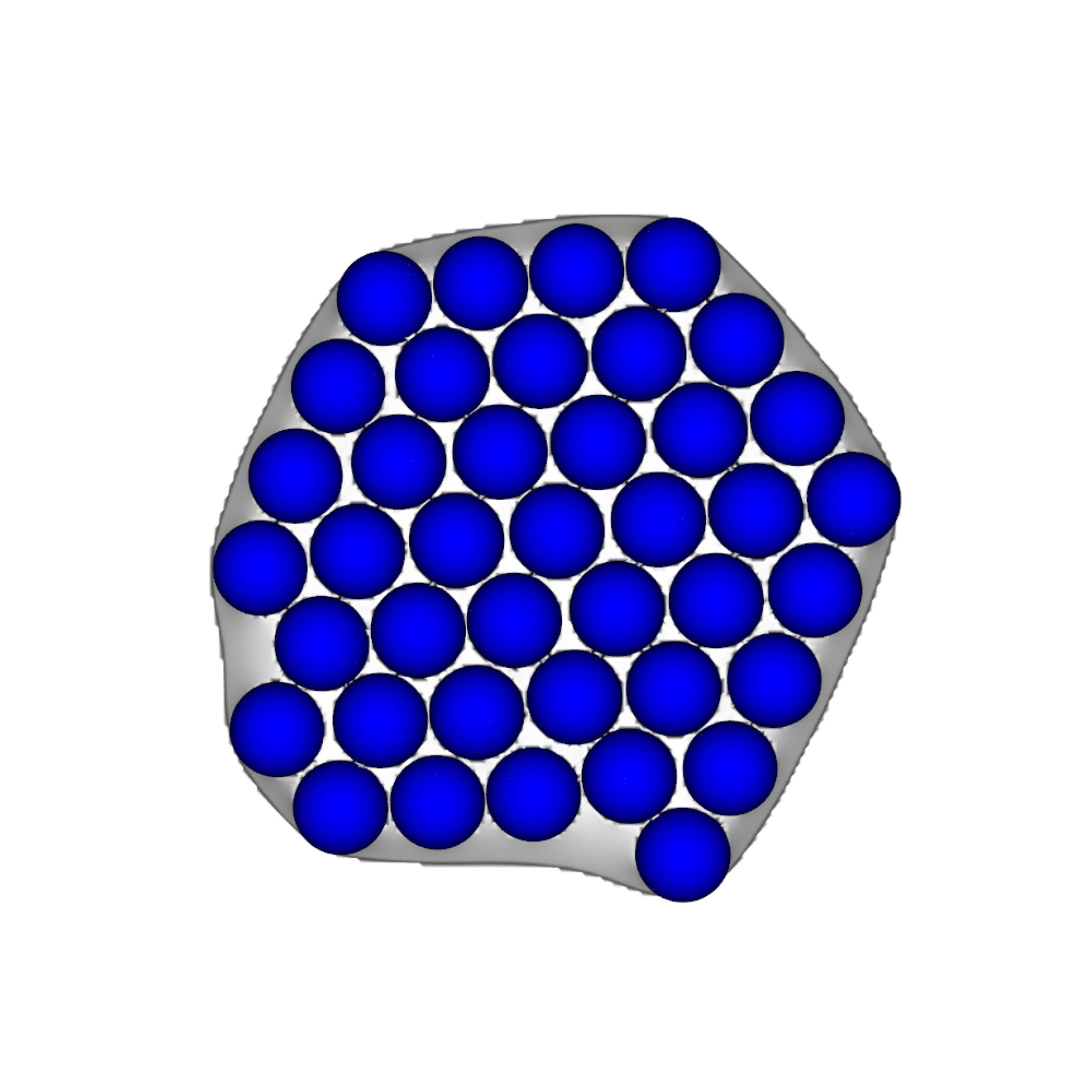}
 \label{fig:p30-zup-3}
 \end{subfigure}
 \begin{subfigure}{.10\textwidth}
\includegraphics[width= 1.0\textwidth]{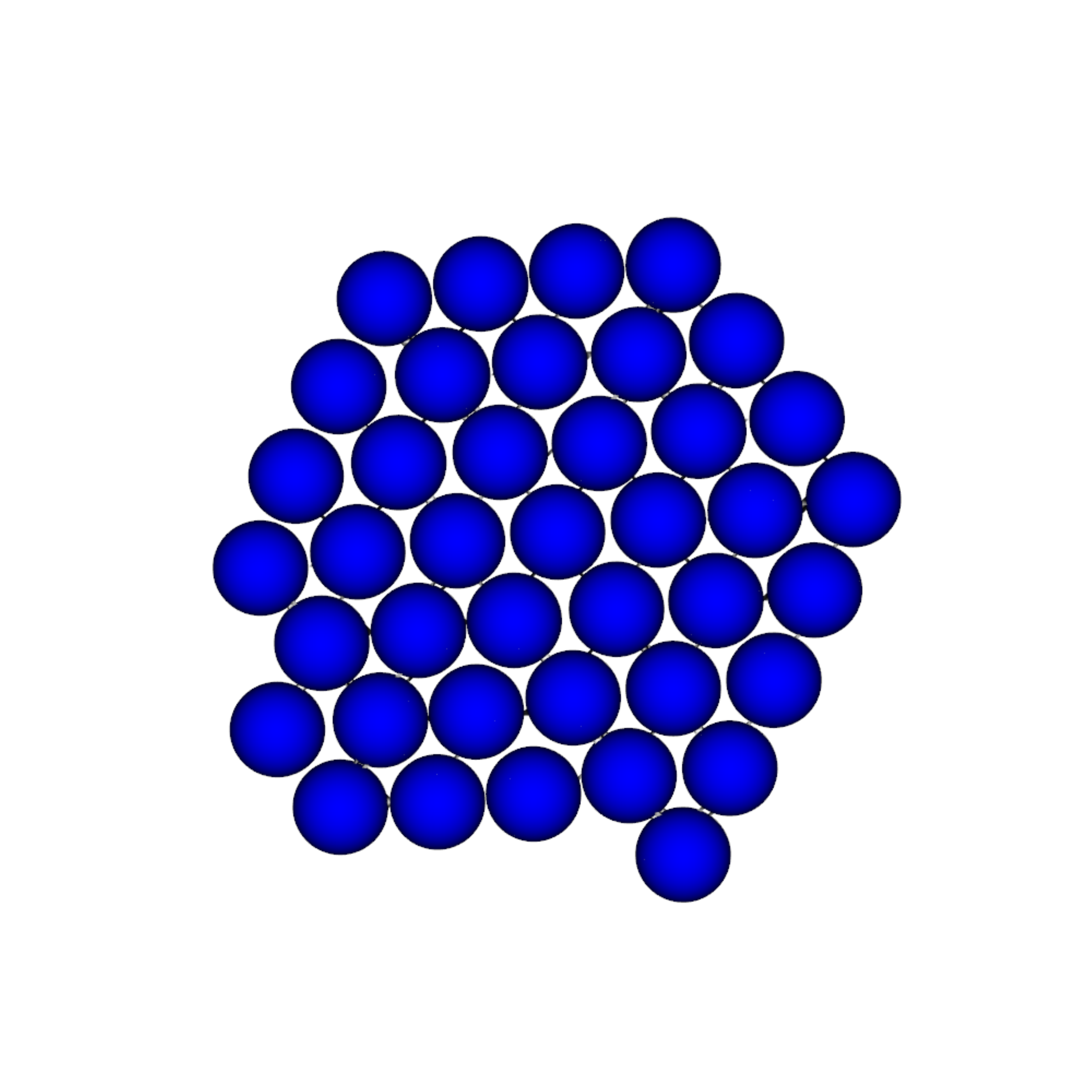}
 \label{fig:p30-zup-4}
\end{subfigure}

\begin{subfigure}{.10\textwidth}
\includegraphics[width= 1.0\textwidth]{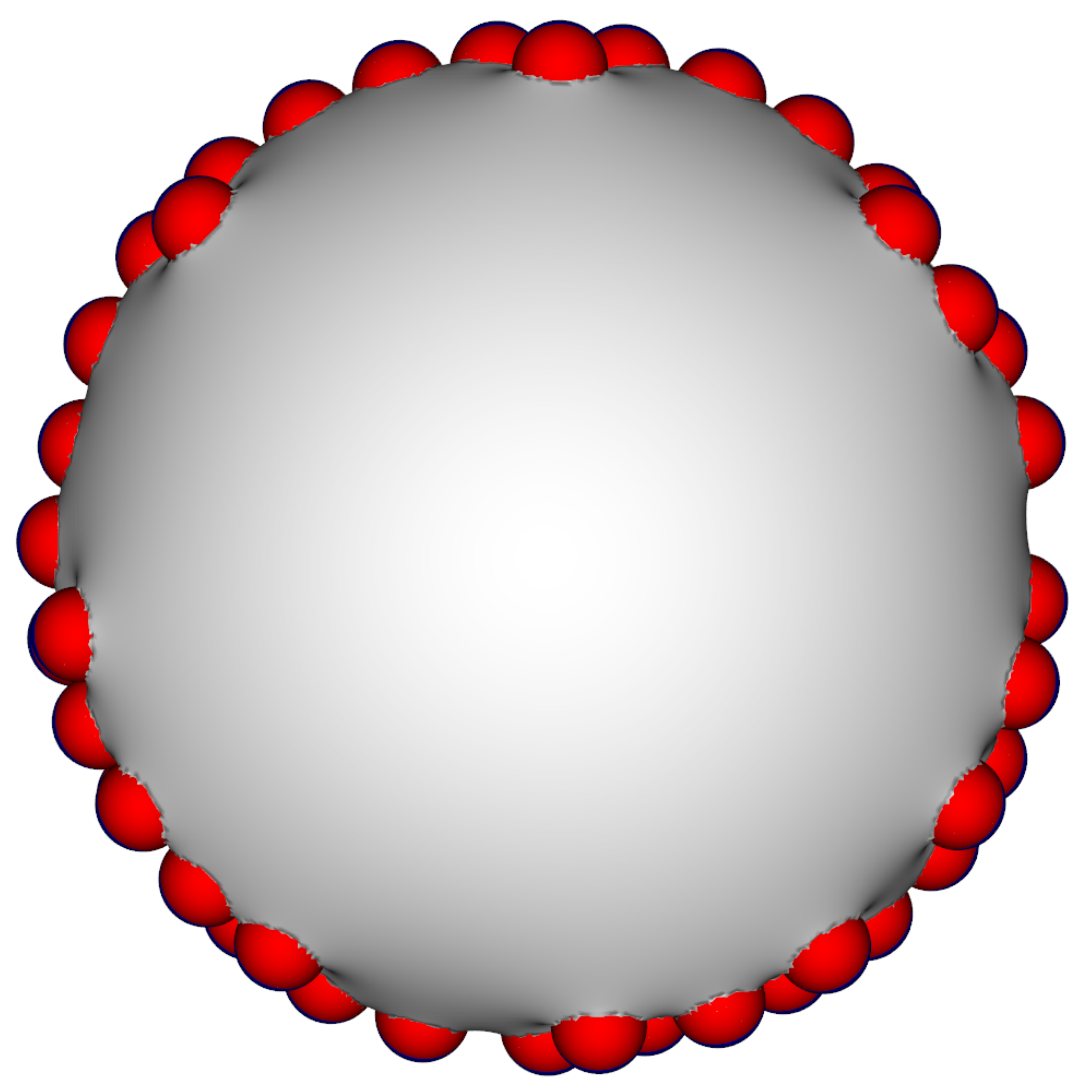}
 \label{fig:p30-zdown-1}
 \end{subfigure}
 \begin{subfigure}{.10\textwidth}
\includegraphics[width= 1.0\textwidth]{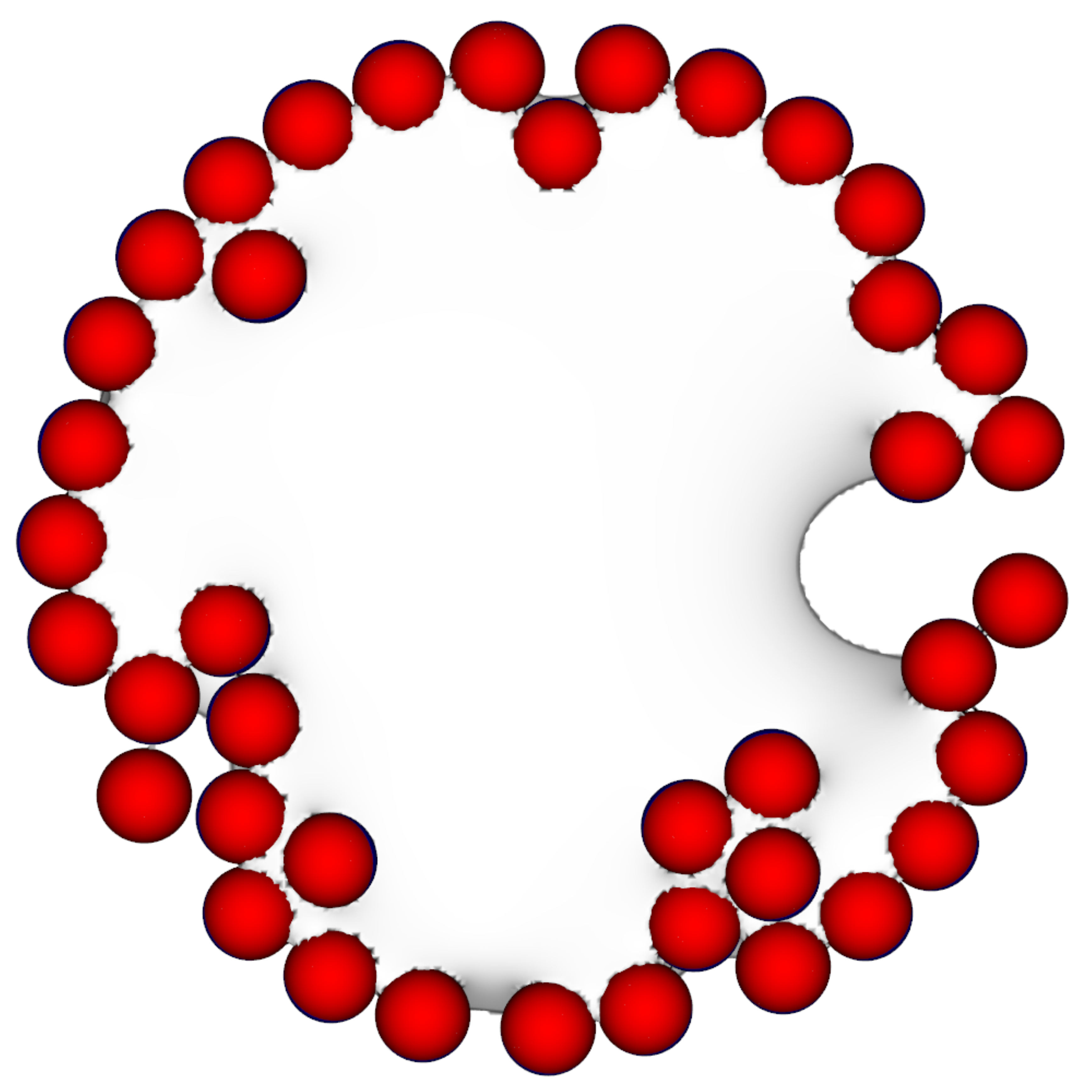}
 \label{fig:p30-zdown-2}
\end{subfigure}
\begin{subfigure}{.1\textwidth}
\includegraphics[width= 1.0\textwidth]{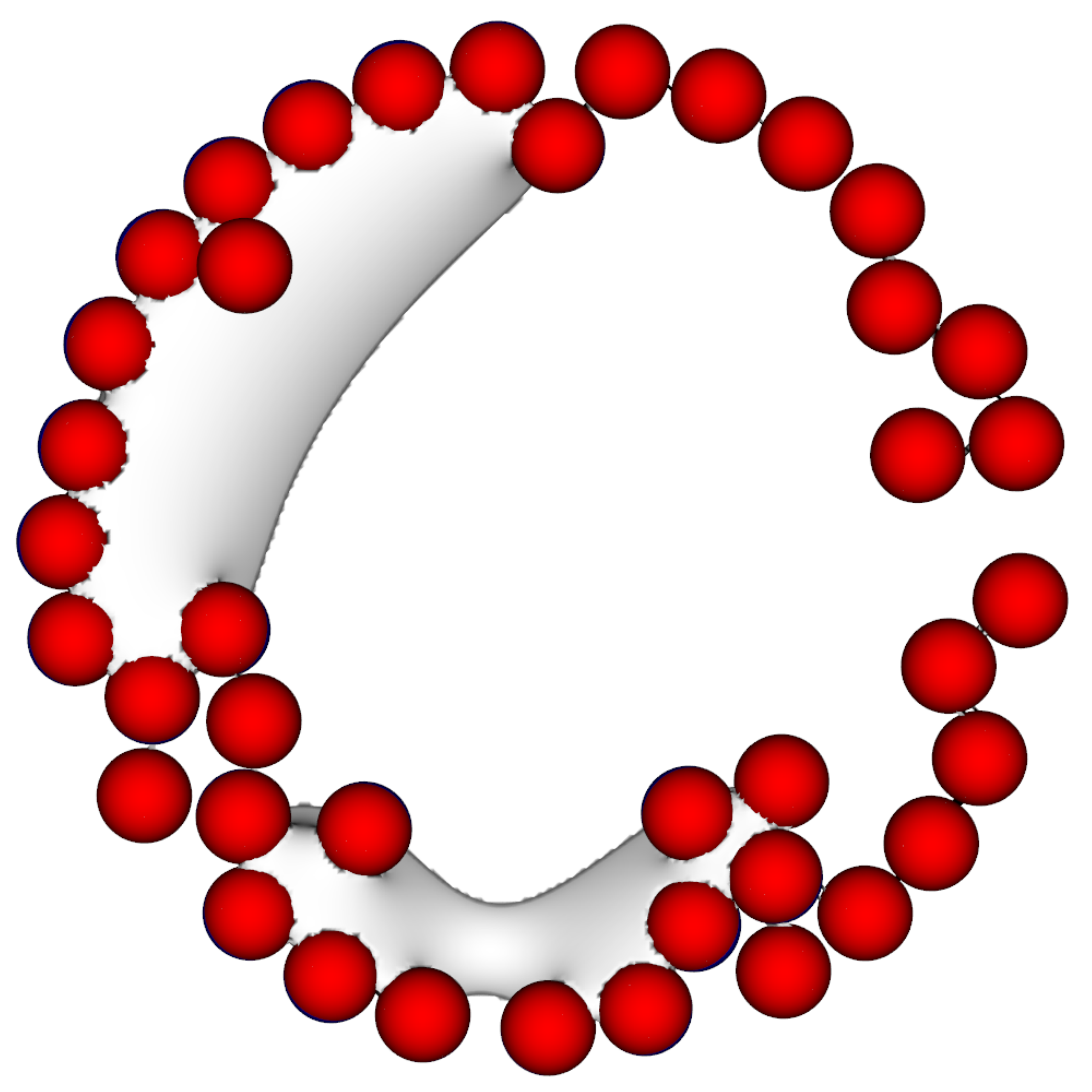}
 \label{fig:p30-zdown-3}
 \end{subfigure}
 \begin{subfigure}{.10\textwidth}
\includegraphics[width= 1.0\textwidth]{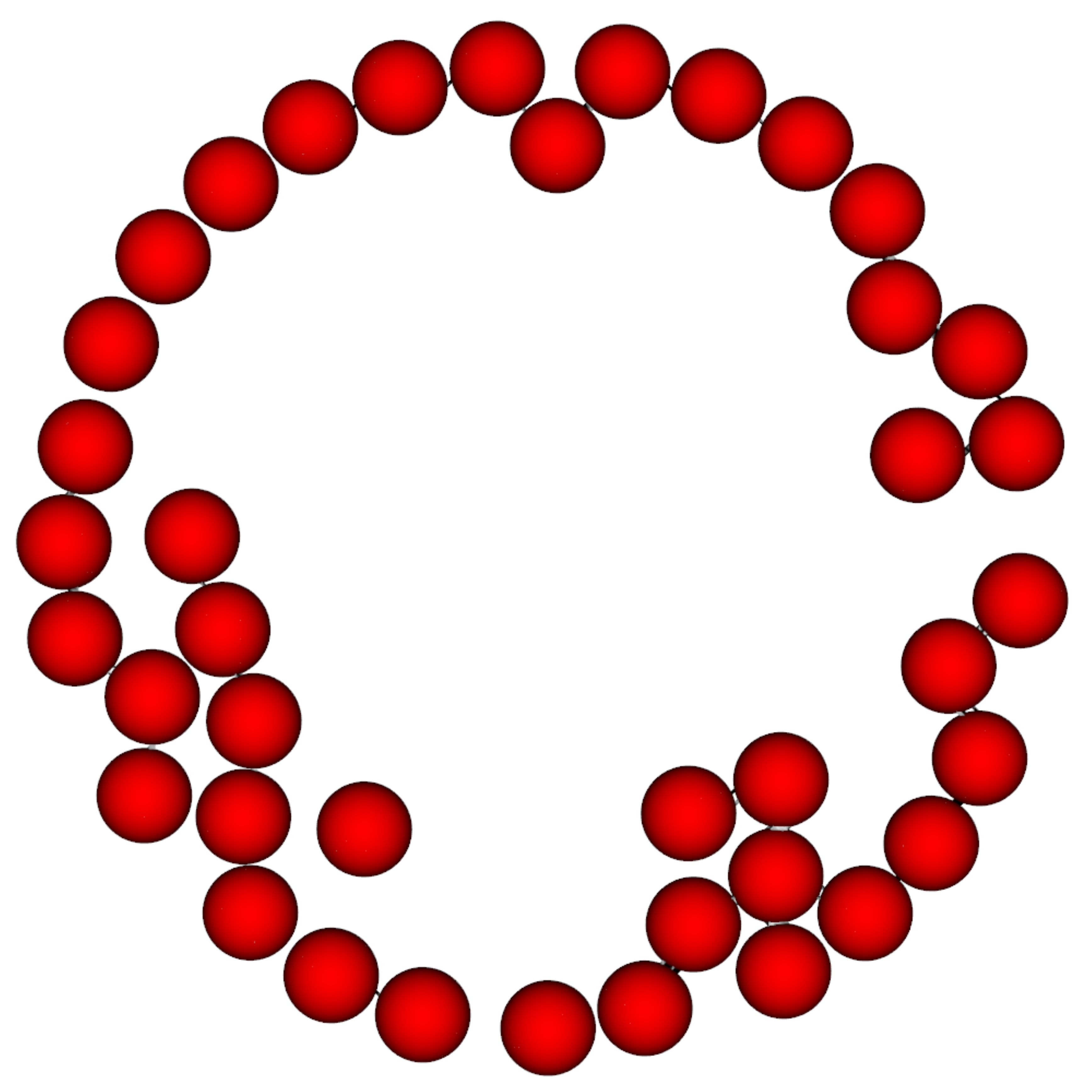}
 \label{fig:p30-zdown-4}
\end{subfigure}

\begin{subfigure}{.10\textwidth}
\includegraphics[width= 1.0\textwidth]{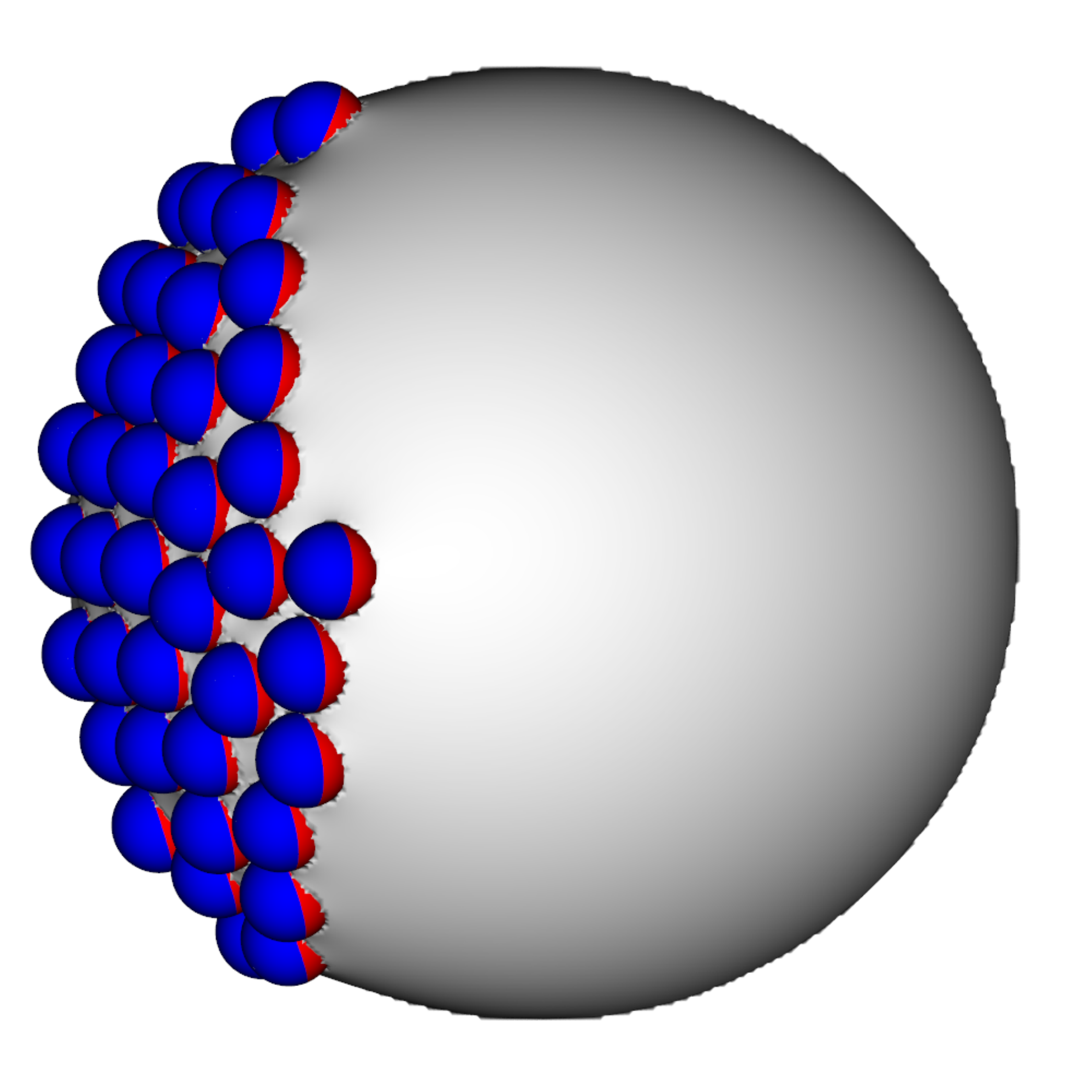}
 \label{fig:p30-zleft-1}
 \end{subfigure}
 \begin{subfigure}{.10\textwidth}
\includegraphics[width= 1.0\textwidth]{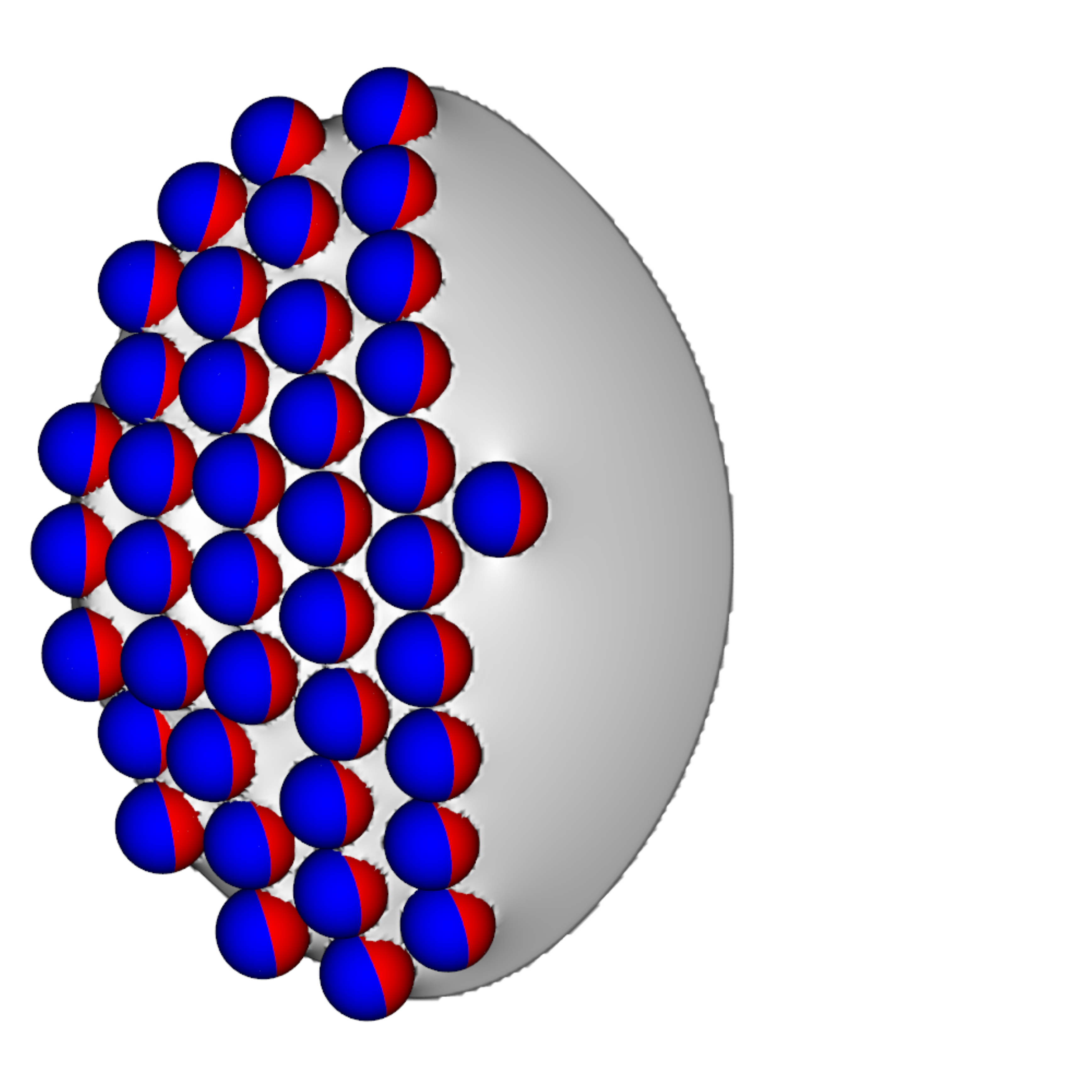}
 \label{fig:p30-zleft-2}
\end{subfigure}
\begin{subfigure}{.1\textwidth}
\includegraphics[width= 1.0\textwidth]{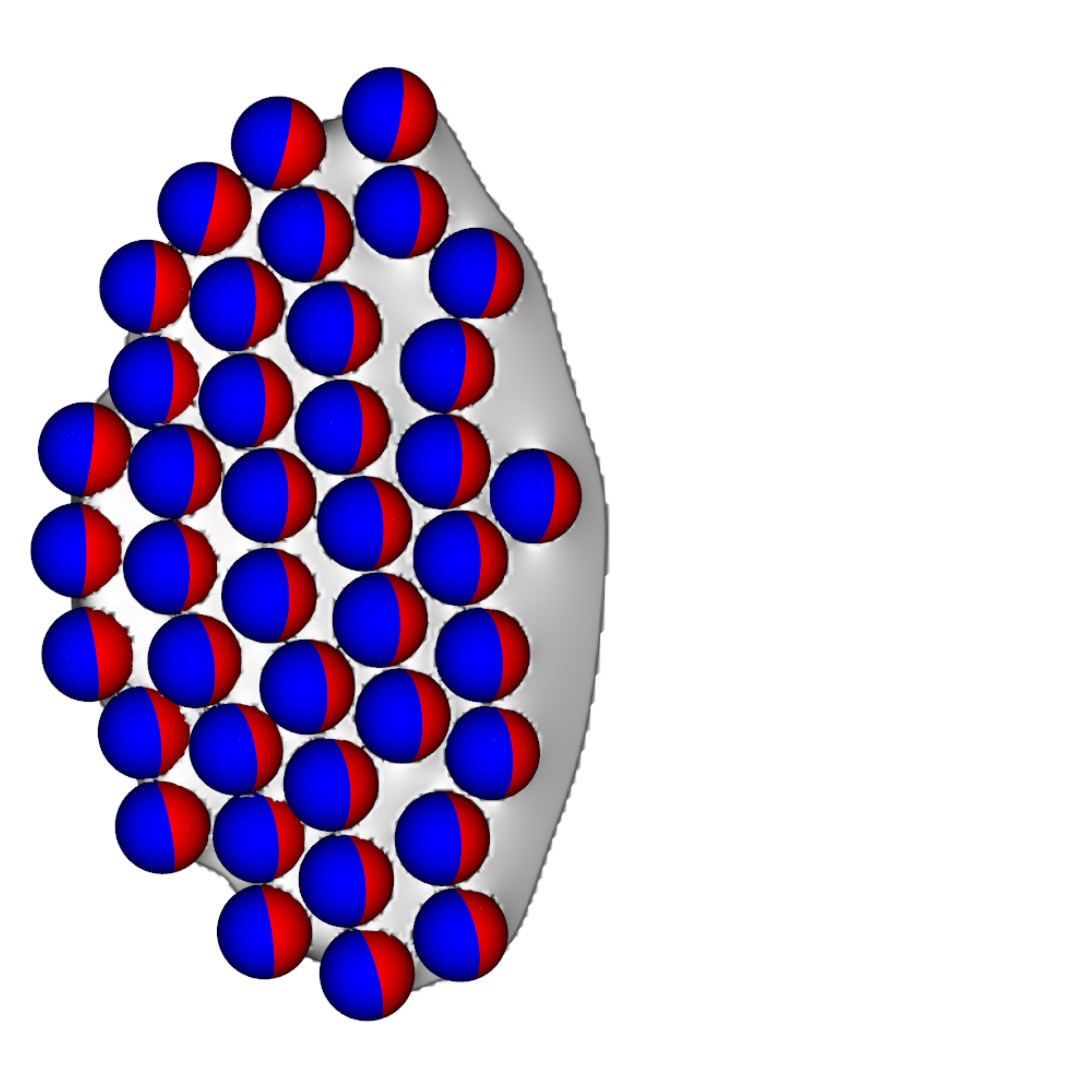}
 \label{fig:p30-zleft-3}
 \end{subfigure}
 \begin{subfigure}{.10\textwidth}
\includegraphics[width= 1.0\textwidth]{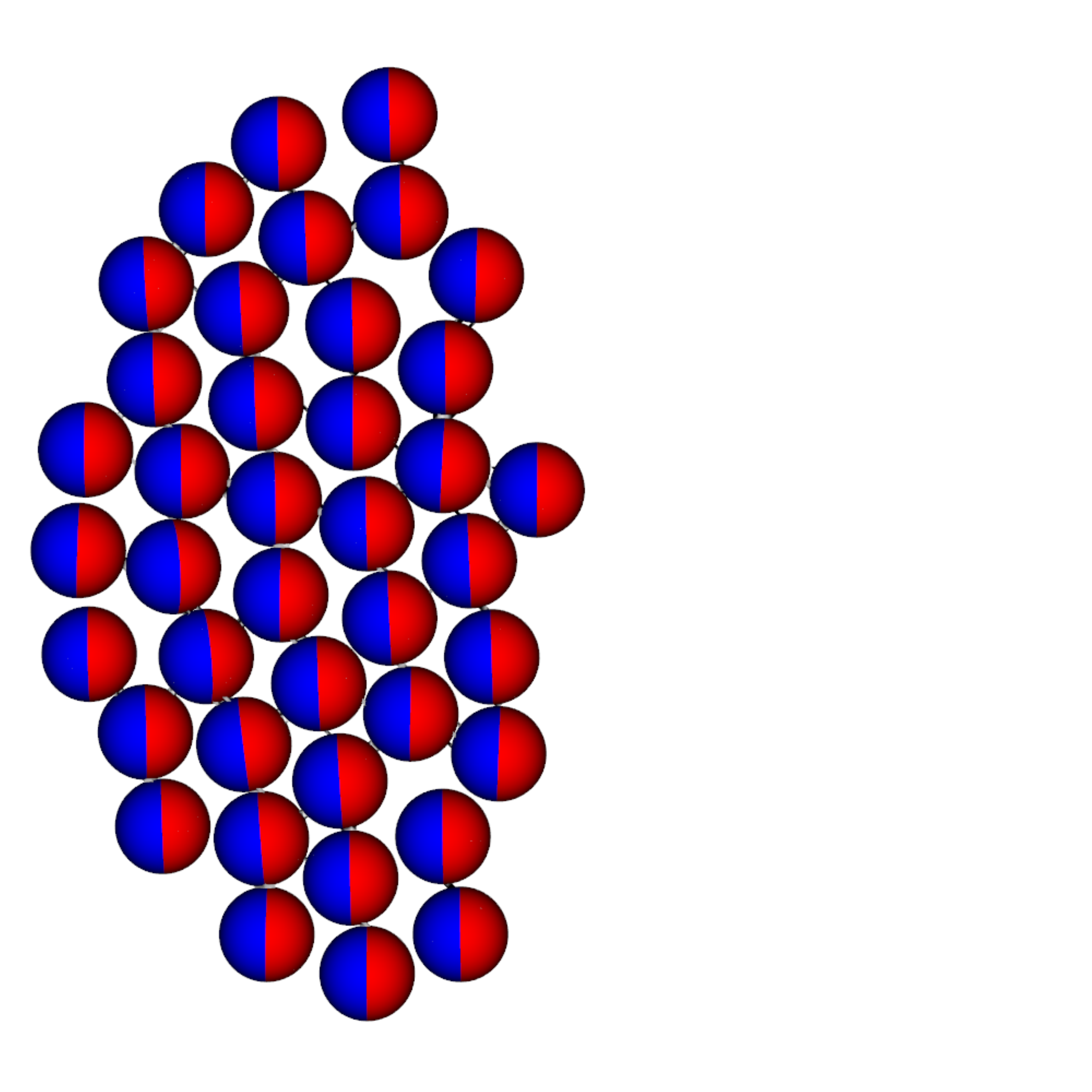}
 \label{fig:p30-zleft-4}
\end{subfigure}

\caption{Snapshots of the assembled structures during droplet evaporation and with an
 upward (a)-(d), downward (e)-(h) and left horizontal (i)-(l) applied magnetic field. Influenced by an upward magnetic field (a-d), the particles remain at the top of the droplet while the contact line decreases before it de-pins. Under a downward magnetic field (e-h), the particles form a ring-like structure that self-pins the contact line, and the contact line deceases continuously. Finally, in a horizontal magnetic field (i-l) the particles migrate in the field direction and cause self-pinning on one side of the droplet, while the contact line recedes on the opposite side and causes inhomogeneous evaporation leading to looser particle arrangements.}
\label{fig:p30-snap}
\end{figure}

When an upward magnetic field is applied, the particles stay at the top during
the evaporation of the droplet (\figref{p30-zup-1}).  The contact angle of the
droplet decreases until it reaches $30^{\circ}$, after which the contact lines
de-pins from the border of the hydrophilic area and moves towards the centre
(\figref{p30-zup-2}).  This is the well-known stick-slip
phenomenon,~\cite{Detlef2015} usually encountered during the evaporation of
droplets on a rough solid substrate.  After the contact line depins, the
droplet continues to dewet (\figref{p30-zup-3}) and approaches the geometry of
a planar film.  The Janus particle will deform the planar film in a monopolar
fashion, introducing attractive capillary forces between particles.  The
attractive capillary forces ensure close packing of the particles, and finally,
a dried highly ordered particle monolayer remains on the substrate
(\figref{p30-zup-4}). [Movie S2]

Under the influence of a downward magnetic field, the particles stay at the
edge of the droplet, forming a ring-like structure during the evaporation of
droplet (\figref{p30-zdown-1}).  The contact angle of the droplet continuously
decreases. We do not observe depinning behaviour once the contact angle reaches
$30^{\circ}$, in contrast with the behaviour observed for an upwardly directed
magnetic field. This is caused by self-pinning due to the particles confined at
the contact line. The colloidal particles experience friction with the
substrate, thus strengthening the pinning of the contact line on the substrate.
When the contact angle decreases further approaching zero degrees, the liquid
film begins to rupture at areas on the edge of the droplet
(\figref{p30-zdown-2}) and the droplet rapidly dries in these areas
(\figref{p30-zdown-3}).  Finally, after full evaporation, a ring-like particle
structure is deposited on the substrate (\figref{p30-zdown-4}). [Movie S3]

Under a left oriented horizontal magnetic field, the particles stay at the left
side of the droplet during the evaporation, as naively expected
(\figref{p30-zleft-1}).  The contact angle of the droplet continuously
decreases, and the contact line depins first at the right side where there is
no self-pinning due to the Janus particles (\figref{p30-zleft-2}). The contact
angle approaches zero degrees and the droplet geometry tends to that of a
planar film, similar to the case of an upwardly directed magnetic field.  In
contrast with the upwardly directed magnetic field, under a horizontal magnetic
field, the Janus particles deform the planar film in a dipolar
fashion~\cite{Xie2015,Xie2016}, and dipolar capillary interactions arise
(\figref{p30-zleft-3}).  This dipolar capillary force is repulsive between
neighbouring particles in the horizontal direction~\cite{Xie2016}, thus the
final structure shows a looser arrangement of particles (\figref{p30-zleft-4})
compared with the particle deposition formed in an upward magnetic field
(\figref{p30-zup-4}). [Movie S4]

Our results demonstrate that the direction of the magnetic field allows one to
tune the particle deposition pattern obtained due to the evaporation and drying
of a Janus particle-laden droplet, which \revisedtext{can find} potential applications in
e.g. advanced printing, electronics or display technology.

%% file: final-janusdrop.tex
\section*{Conclusion}
We investigated the behaviour of magnetic Janus particles at a surface droplet
interface interacting with an external magnetic field and showed that it is
drastically different from the behaviour of such particles adsorbed at a flat
fluid-fluid interface. We found that the particles move to the location where
the particle position vector is parallel to the direction of applied the
field. When multiple Janus particles adsorb at the interface,
the particles arrange hexagonally in contrast to the straight chains observed
for magnetic Janus particles at a flat interface.~\cite{Xie2016} We developed an interface free energy model describing this system, and found that
our model predicts the behaviour observed in our simulations.

Finally, we investigated the behaviour of magnetic Janus particles adsorbed at
a surface droplet interface that undergoes evaporation, finding \revisedtext{interesting} behaviour
at the contact line. In addition, we showed that it is possible to tune the
deposition of the particles by varying the magnetic field direction during
evaporation. 

Our results describe a possible way of creating highly ordered and -- more
importantly -- tunable structures for hierarchical materials assembly. 
Possible applications are ubiquituous not only in the bottom up formulation of
nanostructured surfaces and materials, but also where the ease of
reconfiguration by applying an external magnetic field is of advantage: The
fast response time of magnetic Janus nanoparticles together with their tunable
collective behaviour \revisedtext{have implications} for \revisedtext{advanced} display, E-ink, and sensor
technologies.

%% file: methods.tex
\section*{Methods}
\textbf{Simulation method.}
\label{sec:methods}
We use the lattice Boltzmann method (LBM) for the simulation of fluids. 
The LBM can be treated as an alternative solver for the Navier-Stokes equations in the limit of low Knudsen and Mach numbers and 
has gained popularity due to its straightforward implementation. The locality of the algorithm allows for efficient parallelization.~\cite{Succi2001}
The method's popularity also stems from the variety of multiphase/
multicomponent extensions~\cite{Shan1993,LKLSNJWVH16} and the ability to simulate suspensions.~\cite{ladd-verberg2001}
We utilize the pseudopotential multicomponent LBM model, developed by Shan and
Chen~\cite{Shan1993} and present some relevant details in the following.
Two fluid components are modelled by following the evolution of the
single-particle distribution function discretized in space and time,
\begin{eqnarray}
  \label{eq:LBG}
 f_i^c(\vec{x} + \vec{c}_i \Delta t , t + \Delta t)=f_i^c(\vec{x},t)+\Omega_i^c(\vec{x},t)
  \mbox{,}
\end{eqnarray}
where $i=1,...,19$. $f_i^c(\vec{x},t)$ are the single-particle distribution functions for fluid component $c=1$ or $2$, 
$\vec{c}_i$ is the discrete velocity in the $i$th direction, and 
\begin{equation}
  \label{eq:BGK_collision_operator}
  \Omega_i^c(\vec{x},t) = -\frac{f_i^c(\vec{x},t)- f_i^\mathrm{eq}(\rho^c(\vec{x},t), \vec{u}^c(\vec{x},t))}{\left( \tau^c / \Delta t \right)}
\end{equation}
is the Bhatnagar-Gross-Krook (BGK) collision operator.~\cite{Bhatnagar1954} Here, 
$f_i^\mathrm{eq}(\rho^c(\vec{x},t),\vec{u}^c(\vec{x},t))$ is a
third-order equilibrium distribution function.~\cite{Chen1992}
$\tau^c$ is the relaxation time of component $c$, 
which determines the speed of relaxation of the distribution function towards the equilibrium.
The macroscopic densities and velocities are given as 
$ \rho^c(\vec{x},t) = \rho_0 \sum_if^c_i(\vec{x},t)$, where $\rho_0$ is a reference density,
and $\vec{u}^c(\vec{x},t) = \sum_i  f^c_i(\vec{x},t) \vec{c}_i/\rho^c(\vec{x},t)$, respectively.
For simplicity, We choose the lattice constant $\Delta x$, the timestep $ \Delta t$, 
the unit mass $\rho_0 $ and the relaxation time $\tau^c$ to be unity throughout this paper.

Shan and Chen introduced a pseudopotential interaction between fluid components
$c$ and $c'$. A sufficiently strong interaction parameter ($g^{c\bar{c}}$=0.1
in our case) triggers the separation of components and the formation of a
diffuse interface with a well defined surface tension. The typical width of
this interface is $\approx 5\Delta x$.~\cite{Shan1993} The components separate
into a denser majority phase of density $\rho_{ma}$ and a lighter 
minority phase of density $\rho_{mi}$, respectively. In order to simulate
evaporation, we impose the density of component $c$ at the boundary sites to be
a constant value.~\cite{DennisXieJens2016} In addition, we apply Shan-Chen type
potentials between solid walls and the fluid~\cite{Huang2007} in order to tune
the wettability of the substrate. 

To model the particles, we discretize them on the fluid lattice and couple them 
to the fluid species by means of a modified bounce-back boundary condition as pioneered by Ladd and Aidun.~\cite{ladd-verberg2001, AIDUN1998}
Lubrication forces between particles are properly resolved by the LBM until a distance of one $\Delta x$ between the particle surfaces. If particles get closer, we add a short-range lubrication correction~\cite{ladd-verberg2001} as well as a Hertz force between the particles.~\cite{Jansen2011}
The method was later extended to multiphase flows.~\cite{bib:stratford-adhikari-pagonabarraga-desplat-cates:2005,bib:joshi-sun:2009,Jansen2011}

For particles with magnetic dipole moment $\mathbf{m}$, the dipole-dipole interaction is
\begin{equation}
 U_{ij} = \frac{\mu_0}{4\pi}\frac{m_i m_j}{r_{ij}^3}\left[\mathbf{\hat{o}}_i
\cdot\mathbf{\hat{o}}_j-3(\mathbf{\hat{o}}_i\cdot\mathbf{\hat{r}}_{ij})(\mathbf{\hat{o}}_j\cdot\mathbf{\hat{r}}_{ij})\right],
\end{equation}
where $\mu_0=4\pi \times10^{-7}$ is the magnetic constant, 
$\mathbf{\hat{o}}$ is the unit vector of the orientation of the particle,
$r_{ij}$ is the distance between particle $i$ and $j$,
$\mathbf{\hat{r}}_{ij}=\frac{\mathbf{r}_{i}-\mathbf{r}_{j}}{|\mathbf{r}_{i}-\mathbf{r}_{j}|}$ and
$m_i$ is the magnitude of the magnetic dipole in particle $i$.
The dipole-dipole interaction force is $F_{dd}\approx \frac{\mu_0}{4\pi}\frac{m_i m_j}{r_{ij}^3}$.
A magnetic dipole in an external magnetic field $\mathbf{H}$
experiences a force $F_i=m_i\mathbf{\hat{o}_i} \cdot\Delta \mathbf{H}$,
and a torque $T_i=m_i \mathbf{\hat{o}_i} \times \mathbf{H}$.
In our simulations, we set the dipole moment $|\mathbf{m}|=1$, the magnetic field $|\mathbf{H}|=20$, and 
the particle radius $a=10$ or $a=20$. Thus, the dipole-dipole interaction force 
$F_{dd} \approx \frac{\mu_0}{4\pi}\frac{m_i m_j}{r_{ij}^3}$ is of order $10^{-10}$, and the 
magnetic torque is of order $10$. Based on our previous work, 
we furthermore know that the particle-interface interaction force can be of order $10^{-3}$.~\cite{Xie2016} 
Thus, we can neglect magnetic dipole-dipole interactions between the particles.
Experimentally, the magnetic torque $Me$ can be of order of the capillary torque $\gamma_{12}a^2$. 
Taking $a = 10 nm$, and using the surface tension $\gamma_{12} \approx 70$ $\mathrm{mN/m}$ of 
a water–oil interface, we obtain the magnetic torque
$Me \approx 7\times 10^{-18} Nm$. This torque should be achievable with
a strong magnet (\textit{e.g.,} $H \approx 0.07T$) and a magnetic moment of
$1\times 10^{-16} Nm/T$. The resulting magnetic interaction force is of order $10^{-17} N$, which is much smaller than
the resulting particle-interface interaction force $\gamma_{12} a$ of order $ 10^{-9} N$.
%


For a detailed description of the method and our implementation including,
we refer the reader to the relevant literature.~\cite{Jansen2011,Frijters2012,Xie2015}

We perform simulations of single and multiple particles in a system containing two fluid components. 
The system size is $S = 256 \times 256 \times 144$ and the droplet radius is $100$. Particles
with radius $20$ are used when there are $N=1-6$ particles in the system, while the radius is lowered to $a=10$ for the system containing $N=40$ particles.



\textbf{Interfacial energy of a single Janus particle at a droplet interface.}
When Janus particles are absorbed at the droplet interface, 
the interface can be deformed due to gravity, rotation, anisotropic shape or surface roughness of the particles. 
Our lattice Boltzmann simulations are capable of capturing the interface 
deformations without making any assumptions about the magnitude of the deformations 
or stipulating any particle-fluid boundary conditions, as shown in our 
previous work.~\cite{Xie2015}
To develop a theoretical model to describe the behaviour observed in the simulations, 
we consider particles with a radius much smaller than the capillary length such that we can neglect the effect of gravity.
Our particles
are spheres, thus we can eliminate the deformation due to an
anisotropic shape or a rough surface of the particles.

The interface deformation in our system is mainly induced by the
rotation of the particles.  However, the difficulty in modelling the shape of
the interface and the position of the contact line prevents any exact
analytical expression to exist for this problem.  Park and Lee~\cite{Park2012}
numerically calculate the interface deformation introduced by rotation of
ellipsoidal and dumbbell Janus particles.  They find that the theoretical model
that assumes no interface deformation is able to predict the equilibrium
orientation of nanoparticles reasonably accurately.  Moreover, the interface
deformation is highly dependent on the tilt angle of the Janus particle. The
tilt angle is defined as the angle between particle orientation and normal
direction of the droplet interface.  The interface deformation increases with
an increasing tilt angle for small tilt angles.  In our system, for example,
under an upward magnetic field, the particle will move to the top of the
droplet, which means the tilt angle of the Janus particle with repect to the
normal direction of the droplet interface is zero.  Therefore, we propose a
theoretical model assuming that interface deformation is negligible.

The interface energy of the particle in its equilibrium configuration is
\begin{equation}
 E_{\text{int}} = \gamma_{12}A_{12}^{ \text{int}} + \gamma_{a1}A_{a1}^{\text{int}} +\gamma_{p2}A_{p2}^{\text{int}}
  \mbox{,}
  \end{equation}
where $\gamma_{ij}$ are the interface tensions between phases $i$ and $j$ and $A_{ij}$ are the
contact surface areas between phases $i$ and $j$, where $i,j$ $=$ $\{1$: fluid, $2$: fluid, $a$: apolar, $p$: polar$\}$.
\revisedtext{Under the assumption that the particle radius is substantially smaller than the droplet radius, 
the local interface around the Janus particle can be approximated by a flat plane that is tilted with respect to the horizontal direction by an angle $\phi$. 
Thus, the global curvature of the droplet interface has an effect on the interface energy of this system and we can neglect the effect of the local curvature.} We then obtain $A_{12}^{ \text{int}} = 2\pi R^2-\pi a^2$.
In the case of a symmetric amphiphilic spherical particle, the apolar and polar surface
areas are equal, \revisedtext{\textit{i.e.,}} $A_{a1}=A_{p2} = 2\pi a^2$.

Under a strong magnetic field, the magnetic dipole will align in parallel to the direction of the magnetic field.
At first, we assume the particle's position is fixed, thus the interface energy of a Janus particle can be written as
\begin{eqnarray}
 E_{\text{mag}} = &&\gamma_{12}A_{12}^{\text{mag}} + \gamma_{a1}A_{a1}^{\text{mag}} + \gamma_{p2}A_{p2}^{\text{mag}} \nonumber \\ 
   +&&\gamma_{a2}A_{a2} + \gamma_{p1}A_{p1}
  \mbox{.}
\end{eqnarray}
where $A_{12}^{\text{mag}} =2\pi R^2-\pi a^2$.

The interface energy difference between the magnetic field induced orientation state and the initial state is
\begin{eqnarray}
\!\!\!\!\!\Delta E & = & E_{\text{mag}} - E_{\text{int}} \nonumber \\ 
          & = & \gamma_{12} \left( A_{12}^{\text{mag}} -A_{12}^{ \text{int}} \right) +  \gamma_{a1} \left( A_{a1}^{\text{mag}} - A_{a1}^{\text{int}}  \right) \nonumber \\ 
          & & +  \gamma_{p2}\left(A_{p2}^{\text{mag}} - A_{p2}^{\text{int}} \right) + \gamma_{a2}A_{a2} +\gamma_{p1}A_{p1} 
          \mbox{.}
\end{eqnarray}
Since $ A_{a1}^{\text{int}} = A_{a1}^{\text{mag}} + A_{a2}$ and $ A_{p2}^{\text{int}} = A_{p2}^{\text{mag}} +A_{p1} $, we obtain 
\begin{eqnarray}
 \Delta E  &=& \gamma_{12}(A_{12}^{\text{mag}} -A_{12}^{\text{int}})  +(\gamma_{a2}-\gamma_{a1}) A_{a2} \nonumber \\ 
         & & + (\gamma_{p1}-\gamma_{p2})A_{p1}
   \mbox{,}
 \label{eq:free}
\end{eqnarray}

According to Young's boundary conditions,~\cite{Ondurcuhu1990} we obtain
\begin{equation}
 \cos \theta_A = \frac{\gamma_{a1}-\gamma_{a2}} { \gamma_{12}} \mbox{, } \qquad \cos \theta_P = \frac{\gamma_{p1}-\gamma_{p2}} {\gamma_{12}}
 \mbox{.}
\end{equation}
Using $\cos \theta_A = -\cos \theta_P = -\sin \beta$ resulting from
the condition that the two hemispheres have opposite wettabilities, we can further simply \eqnref{free} to
\begin{equation}
  \Delta E = \gamma_{12} (A_{12}^{\text{mag}} -A_{12}^{\text{int}})  +  \gamma_{12} (A_{a2} + A_{p1}) \sin \beta 
    \mbox{.}
 \label{eq:free_easy}
\end{equation}
Under the assumption that the interface deformation is negligible, $A_{12}^{\text{mag}} =A_{12}^{\text{int}}$, $A_{a2} =A_{p1}= \frac{\phi}{2\pi} 4\pi a^2= 2\phi a^2 $. 
Therefore, \eqnref{free} finally reduces to
\begin{equation}
\Delta E = 4 \phi a^2 \gamma_{12} \sin \beta 
\mbox{.}
 \label{eq:eflat}
\end{equation}

\textbf{Total interface energy of multiple particles at a droplet interface.}
If there are $N$ particles adsorbed at the interface, 
the total interface energy becomes
\begin{equation}
 \Delta E_T = 4  a^2 \gamma_{12} \sin \beta \sum_{i}^{N} \phi_i.
 \label{eq:totalenergy}
\end{equation}
Based on a geometrical analysis, we get 
\begin{equation}
 \sin \phi/2 = \frac{|\mathbf{r}_i - \mathbf{r}_0|}{2R}
 \mbox{.}
\end{equation}
Under the effect of an upward magnetic field, the particles move to the top, resulting a small $\phi$ angle.
Therefore, we can write 
\begin{equation}
 \phi/2 \approx \frac{|\mathbf{r}_i - \mathbf{r}_0|}{2R}
 \mbox{.}
\end{equation}
Then, we can rewrite \eqnref{totalenergy} into
\begin{equation}
 \Delta E_T = \frac {4 a^2 \gamma_{12} \sin \beta }{R} \sum_{i}^{N} |\mathbf{r_i} - \mathbf{r_0}|. 
 \label{eq:totalenergy2}
\end{equation}

%% file: ref-acsnano.bbl
\providecommand*\mcitethebibliography{\thebibliography}
\csname @ifundefined\endcsname{endmcitethebibliography}
  {\let\endmcitethebibliography\endthebibliography}{}

%% file: main-janusdrop.bbl
\begin{thebibliography}{0}%
\makeatletter
\providecommand \@ifxundefined [1]{%
 \@ifx{#1\undefined}
}%
\providecommand \@ifnum [1]{%
 \ifnum #1\expandafter \@firstoftwo
 \else \expandafter \@secondoftwo
 \fi
}%
\providecommand \@ifx [1]{%
 \ifx #1\expandafter \@firstoftwo
 \else \expandafter \@secondoftwo
 \fi
}%
\providecommand \natexlab [1]{#1}%
\providecommand \enquote  [1]{``#1''}%
\providecommand \bibnamefont  [1]{#1}%
\providecommand \bibfnamefont [1]{#1}%
\providecommand \citenamefont [1]{#1}%
\providecommand \href@noop [0]{\@secondoftwo}%
\providecommand \href [0]{\begingroup \@sanitize@url \@href}%
\providecommand \@href[1]{\@@startlink{#1}\@@href}%
\providecommand \@@href[1]{\endgroup#1\@@endlink}%
\providecommand \@sanitize@url [0]{\catcode `\\12\catcode `\$12\catcode
  `\&12\catcode `\#12\catcode `\^12\catcode `\_12\catcode `\%12\relax}%
\providecommand \@@startlink[1]{}%
\providecommand \@@endlink[0]{}%
\providecommand \url  [0]{\begingroup\@sanitize@url \@url }%
\providecommand \@url [1]{\endgroup\@href {#1}{\urlprefix }}%
\providecommand \urlprefix  [0]{URL }%
\providecommand \Eprint [0]{\href }%
\providecommand \doibase [0]{http://dx.doi.org/}%
\providecommand \selectlanguage [0]{\@gobble}%
\providecommand \bibinfo  [0]{\@secondoftwo}%
\providecommand \bibfield  [0]{\@secondoftwo}%
\providecommand \translation [1]{[#1]}%
\providecommand \BibitemOpen [0]{}%
\providecommand \bibitemStop [0]{}%
\providecommand \bibitemNoStop [0]{.\EOS\space}%
\providecommand \EOS [0]{\spacefactor3000\relax}%
\providecommand \BibitemShut  [1]{\csname bibitem#1\endcsname}%
\let\auto@bib@innerbib\@empty
\end{thebibliography}%


\begin{mcitethebibliography}{42}
\providecommand*\natexlab[1]{#1}
\providecommand*\mciteSetBstSublistMode[1]{}
\providecommand*\mciteSetBstMaxWidthForm[2]{}
\providecommand*\mciteBstWouldAddEndPuncttrue
  {\def\EndOfBibitem{\unskip.}}
\providecommand*\mciteBstWouldAddEndPunctfalse
  {\let\EndOfBibitem\relax}
\providecommand*\mciteSetBstMidEndSepPunct[3]{}
\providecommand*\mciteSetBstSublistLabelBeginEnd[3]{}
\providecommand*\EndOfBibitem{}
\mciteSetBstSublistMode{f}
\mciteSetBstMaxWidthForm{subitem}{(\alph{mcitesubitemcount})}
\mciteSetBstSublistLabelBeginEnd
  {\mcitemaxwidthsubitemform\space}
  {\relax}
  {\relax}

\bibitem[Boal et~al.(2000)Boal, Ilhan, De~Rouchey, Thurn-Albrecht, Russell, and
  Rotello]{Boal2000}
Boal,~A.; Ilhan,~F.; De~Rouchey,~J.; Thurn-Albrecht,~T.; Russell,~T.;
  Rotello,~V. {Self-Assembly of Nanoparticles into Structured Spherical and
  Network Aggregates}. \emph{Nature} \textbf{2000}, \emph{404}, 746--748\relax
\mciteBstWouldAddEndPuncttrue
\mciteSetBstMidEndSepPunct{\mcitedefaultmidpunct}
{\mcitedefaultendpunct}{\mcitedefaultseppunct}\relax
\EndOfBibitem
\bibitem[Olson et~al.(2009)Olson, Coskun, Klajn, Fang, Dey, Browne, Grzybowski,
  and Stoddart]{Olson2009}
Olson,~M.~A.; Coskun,~A.; Klajn,~R.; Fang,~L.; Dey,~S.~K.; Browne,~K.~P.;
  Grzybowski,~B.~A.; Stoddart,~J.~F. {Assembly of Polygonal Nanoparticle
  Clusters Directed by Reversible Noncovalent Bonding Interactions}. \emph{Nano
  Lett.} \textbf{2009}, \emph{9}, 3185--3190\relax
\mciteBstWouldAddEndPuncttrue
\mciteSetBstMidEndSepPunct{\mcitedefaultmidpunct}
{\mcitedefaultendpunct}{\mcitedefaultseppunct}\relax
\EndOfBibitem
\bibitem[Hermanson et~al.(2001)Hermanson, Lumsdon, Williams, Kaler, and
  Velev]{Hermanson2001}
Hermanson,~K.~D.; Lumsdon,~S.~O.; Williams,~J.~P.; Kaler,~E.~W.; Velev,~O.~D.
  {Dielectrophoretic Assembly of Electrically Functional Microwires from
  Nanoparticle Suspensions.} \emph{Science} \textbf{2001}, \emph{294},
  1082--1086\relax
\mciteBstWouldAddEndPuncttrue
\mciteSetBstMidEndSepPunct{\mcitedefaultmidpunct}
{\mcitedefaultendpunct}{\mcitedefaultseppunct}\relax
\EndOfBibitem
\bibitem[Weiss(2008)]{Weiss2008}
Weiss,~P.~S. {Hierarchical Assembly}. \emph{ACS Nano} \textbf{2008}, \emph{2},
  1085--1087\relax
\mciteBstWouldAddEndPuncttrue
\mciteSetBstMidEndSepPunct{\mcitedefaultmidpunct}
{\mcitedefaultendpunct}{\mcitedefaultseppunct}\relax
\EndOfBibitem
\bibitem[Grzelczak et~al.(2010)Grzelczak, Vermant, Furst, and
  Liz-Marz{\'{a}}n]{Grzelczak2010}
Grzelczak,~M.; Vermant,~J.; Furst,~E.~M.; Liz-Marz{\'{a}}n,~L.~M. {Directed
  Self-Assembly of Nanoparticles}. \emph{ACS Nano} \textbf{2010}, \emph{4},
  3591--605\relax
\mciteBstWouldAddEndPuncttrue
\mciteSetBstMidEndSepPunct{\mcitedefaultmidpunct}
{\mcitedefaultendpunct}{\mcitedefaultseppunct}\relax
\EndOfBibitem
\bibitem[Binks and Fletcher(2001)Binks, and Fletcher]{Binks2001}
Binks,~B.~P.; Fletcher,~P. D.~I. {Particles adsorbed at the Oil-Water
  Interface: A Theoretical Comparison between Spheres of Uniform Wettability
  and “Janus” Particles}. \emph{Langmuir} \textbf{2001}, \emph{17},
  4708--4710\relax
\mciteBstWouldAddEndPuncttrue
\mciteSetBstMidEndSepPunct{\mcitedefaultmidpunct}
{\mcitedefaultendpunct}{\mcitedefaultseppunct}\relax
\EndOfBibitem
\bibitem[Lin et~al.(2005)Lin, B{\"{o}}ker, Skaff, Cookson, Dinsmore, Emrick,
  and Russell]{Lin2005}
Lin,~Y.; B{\"{o}}ker,~A.; Skaff,~H.; Cookson,~D.; Dinsmore,~D.; Emrick,~T.;
  Russell,~T.~P. {Nanoparticle Assembly at Fluid Interfaces: Structure and
  Dynamics}. \emph{Langmuir} \textbf{2005}, \emph{21}, 191--194\relax
\mciteBstWouldAddEndPuncttrue
\mciteSetBstMidEndSepPunct{\mcitedefaultmidpunct}
{\mcitedefaultendpunct}{\mcitedefaultseppunct}\relax
\EndOfBibitem
\bibitem[B{\"{o}}ker et~al.(2007)B{\"{o}}ker, He, Emrick, and
  Russell]{Boker2007}
B{\"{o}}ker,~A.; He,~J.; Emrick,~T.; Russell,~T.~P. {Self-Assembly of
  Nanoparticles at Interfaces}. \emph{Soft Matter} \textbf{2007}, \emph{3},
  1231--1248\relax
\mciteBstWouldAddEndPuncttrue
\mciteSetBstMidEndSepPunct{\mcitedefaultmidpunct}
{\mcitedefaultendpunct}{\mcitedefaultseppunct}\relax
\EndOfBibitem
\bibitem[Jiang et~al.(2011)Jiang, Zeng, Chen, and Yu]{Jiang2011}
Jiang,~X.; Zeng,~Q.; Chen,~C.; Yu,~A. {Self-Assembly of Particles: Some
  Thoughts and Comments}. \emph{J. Mater. Chem.} \textbf{2011}, \emph{21},
  16797--16805\relax
\mciteBstWouldAddEndPuncttrue
\mciteSetBstMidEndSepPunct{\mcitedefaultmidpunct}
{\mcitedefaultendpunct}{\mcitedefaultseppunct}\relax
\EndOfBibitem
\bibitem[Park and Lee(2012)Park, and Lee]{Park2012}
Park,~B.~J.; Lee,~D. {Equilibrium Orientation of Nonspherical Janus Particles
  at Fluid-Fluid Interfaces}. \emph{ACS Nano} \textbf{2012}, \emph{6},
  782--790\relax
\mciteBstWouldAddEndPuncttrue
\mciteSetBstMidEndSepPunct{\mcitedefaultmidpunct}
{\mcitedefaultendpunct}{\mcitedefaultseppunct}\relax
\EndOfBibitem
\bibitem[Davies et~al.(2014)Davies, Kr\"{u}ger, Coveney, Harting, and
  Bresme]{Gary2014a}
Davies,~G.~B.; Kr\"{u}ger,~T.; Coveney,~P.~V.; Harting,~J.; Bresme,~F.
  {Interface Deformations Affect the Orientation Transition of Magnetic
  Ellipsoidal Particles Adsorbed at Fluid-Fluid Interfaces.} \emph{Soft Matter}
  \textbf{2014}, \emph{10}, 6742--6748\relax
\mciteBstWouldAddEndPuncttrue
\mciteSetBstMidEndSepPunct{\mcitedefaultmidpunct}
{\mcitedefaultendpunct}{\mcitedefaultseppunct}\relax
\EndOfBibitem
\bibitem[Xie et~al.(2015)Xie, Davies, G{\"u}nther, and Harting]{Xie2015}
Xie,~Q.; Davies,~G.~B.; G{\"u}nther,~F.; Harting,~J. {{Tunable} Dipolar
  Capillary Deformations for Magnetic {Janus} Particles at Fluid-Fluid
  Interfaces}. \emph{Soft Matter} \textbf{2015}, \emph{11}, 3581--3588\relax
\mciteBstWouldAddEndPuncttrue
\mciteSetBstMidEndSepPunct{\mcitedefaultmidpunct}
{\mcitedefaultendpunct}{\mcitedefaultseppunct}\relax
\EndOfBibitem
\bibitem[Newton and Buzza(2016)Newton, and Buzza]{Martin2016}
Newton,~B.~J.; Buzza,~D. M.~A. {Magnetic Cylindrical Colloids at Liquid
  Interfaces Exhibit Non-Volatile Switching of Their Orientation in an External
  Field}. \emph{Soft Matter} \textbf{2016}, \emph{12}, 5285--5296\relax
\mciteBstWouldAddEndPuncttrue
\mciteSetBstMidEndSepPunct{\mcitedefaultmidpunct}
{\mcitedefaultendpunct}{\mcitedefaultseppunct}\relax
\EndOfBibitem
\bibitem[Davies et~al.(2014)Davies, Kr\"{u}ger, Coveney, Harting, and
  Bresme]{Gary2014b}
Davies,~G.~B.; Kr\"{u}ger,~T.; Coveney,~P.~V.; Harting,~J.; Bresme,~F.
  {Assembling Ellipsoidal Particles at Fluid Interfaces Using Switchable
  Dipolar Capillary Interactions}. \emph{Adv. Mater.} \textbf{2014}, \emph{26},
  6715--6719\relax
\mciteBstWouldAddEndPuncttrue
\mciteSetBstMidEndSepPunct{\mcitedefaultmidpunct}
{\mcitedefaultendpunct}{\mcitedefaultseppunct}\relax
\EndOfBibitem
\bibitem[Xie et~al.(2016)Xie, Davies, and Harting]{Xie2016}
Xie,~Q.; Davies,~G.~B.; Harting,~J. {{Controlled} Capillary Assembly of
  Magnetic {Janus} Particles at Fluid-Fluid Interfaces}. \emph{Soft Matter}
  \textbf{2016}, \emph{12}, 6566--6574\relax
\mciteBstWouldAddEndPuncttrue
\mciteSetBstMidEndSepPunct{\mcitedefaultmidpunct}
{\mcitedefaultendpunct}{\mcitedefaultseppunct}\relax
\EndOfBibitem
\bibitem[Cavallaro et~al.(2011)Cavallaro, Botto, Lewandowski, Wang, and
  Stebe]{Cavallaro2011}
Cavallaro,~M.; Botto,~L.; Lewandowski,~E.~P.; Wang,~M.; Stebe,~K.~J.
  {Curvature-Driven Capillary Migration and Assembly of Rod-Like Particles}.
  \emph{Proc. Natl. Acad. Sci. U. S. A.} \textbf{2011}, \emph{108},
  20923--20928\relax
\mciteBstWouldAddEndPuncttrue
\mciteSetBstMidEndSepPunct{\mcitedefaultmidpunct}
{\mcitedefaultendpunct}{\mcitedefaultseppunct}\relax
\EndOfBibitem
\bibitem[Ershov et~al.(2013)Ershov, Sprakel, Appel, {Cohen Stuart}, and van~der
  Gucht]{Ershov2013}
Ershov,~D.; Sprakel,~J.; Appel,~J.; {Cohen Stuart},~M.; van~der Gucht,~J.
  {Capillarity-Induced Ordering of Spherical Colloids on an Interface with
  Anisotropic Curvature.} \emph{Proc. Natl. Acad. Sci. U. S. A.} \textbf{2013},
  \emph{110}, 9220--9224\relax
\mciteBstWouldAddEndPuncttrue
\mciteSetBstMidEndSepPunct{\mcitedefaultmidpunct}
{\mcitedefaultendpunct}{\mcitedefaultseppunct}\relax
\EndOfBibitem
\bibitem[Hessling et~al.(2017)Hessling, Xie, and Harting]{DennisXieJens2016}
Hessling,~D.; Xie,~Q.; Harting,~J. {{Diffusion} Dominated Evaporation in
  Multicomponent Lattice {B}oltzmann Simulations}. \emph{J. Chem. Phys.}
  \textbf{2017}, \emph{146}, 054111\relax
\mciteBstWouldAddEndPuncttrue
\mciteSetBstMidEndSepPunct{\mcitedefaultmidpunct}
{\mcitedefaultendpunct}{\mcitedefaultseppunct}\relax
\EndOfBibitem
\bibitem[Jansen and Harting(2011)Jansen, and Harting]{Jansen2011}
Jansen,~F.; Harting,~J. {From Bijels to Pickering Emulsions: A Lattice
  Boltzmann Study}. \emph{Phys. Rev. E} \textbf{2011}, \emph{83}, 046707\relax
\mciteBstWouldAddEndPuncttrue
\mciteSetBstMidEndSepPunct{\mcitedefaultmidpunct}
{\mcitedefaultendpunct}{\mcitedefaultseppunct}\relax
\EndOfBibitem
\bibitem[Frijters et~al.(2012)Frijters, G\"{u}nther, and Harting]{Frijters2012}
Frijters,~S.; G\"{u}nther,~F.; Harting,~J. {Effects of Nanoparticles and
  Surfactant on Droplets in Shear Flow}. \emph{Soft Matter} \textbf{2012},
  \emph{8}, 6542--6556\relax
\mciteBstWouldAddEndPuncttrue
\mciteSetBstMidEndSepPunct{\mcitedefaultmidpunct}
{\mcitedefaultendpunct}{\mcitedefaultseppunct}\relax
\EndOfBibitem
\bibitem[Kr\"uger et~al.(2013)Kr\"uger, Frijters, G\"unther, Kaoui, and
  Harting]{KFGKH13}
Kr\"uger,~T.; Frijters,~S.; G\"unther,~F.; Kaoui,~B.; Harting,~J. {Numerical
  Simulations of Complex Fluid-Fluid Interface Dynamics}. \emph{Eur. Phys. J.:
  Spec. Top.} \textbf{2013}, \emph{222}, 177--198\relax
\mciteBstWouldAddEndPuncttrue
\mciteSetBstMidEndSepPunct{\mcitedefaultmidpunct}
{\mcitedefaultendpunct}{\mcitedefaultseppunct}\relax
\EndOfBibitem
\bibitem[Duan et~al.(2004)Duan, Wang, Kurth, and M{\"{o}}hwald]{Duan2004}
Duan,~H.; Wang,~D.; Kurth,~D.~G.; M{\"{o}}hwald,~H. {Directing Self-Assembly of
  Nanoparticles at Water/Oil Interfaces}. \emph{Angew. Chem., Int. Ed.}
  \textbf{2004}, \emph{43}, 5639--5642\relax
\mciteBstWouldAddEndPuncttrue
\mciteSetBstMidEndSepPunct{\mcitedefaultmidpunct}
{\mcitedefaultendpunct}{\mcitedefaultseppunct}\relax
\EndOfBibitem
\bibitem[Lu et~al.(2014)Lu, Aabdin, Loh, Bhattacharya, and Mirsaidov]{Lu2014}
Lu,~J.; Aabdin,~Z.; Loh,~N.~D.; Bhattacharya,~D.; Mirsaidov,~U. {Nanoparticle
  Dynamics in a Nanodroplet}. \emph{Nano Lett.} \textbf{2014}, \emph{14},
  2111--2115\relax
\mciteBstWouldAddEndPuncttrue
\mciteSetBstMidEndSepPunct{\mcitedefaultmidpunct}
{\mcitedefaultendpunct}{\mcitedefaultseppunct}\relax
\EndOfBibitem
\bibitem[Bai et~al.(2014)Bai, Hayes, Jin, Shui, {Chuan Yi}, Wang, Zhang, and
  Zhou]{Bai2014}
Bai,~P.; Hayes,~R.; Jin,~M.; Shui,~L.; {Chuan Yi},~Z.; Wang,~L.; Zhang,~X.;
  Zhou,~G.~F. {Review of Paper-Like Display Technologies}. \emph{Prog.
  Electromagn. Res.} \textbf{2014}, \emph{147}, 95--116\relax
\mciteBstWouldAddEndPuncttrue
\mciteSetBstMidEndSepPunct{\mcitedefaultmidpunct}
{\mcitedefaultendpunct}{\mcitedefaultseppunct}\relax
\EndOfBibitem
\bibitem[Lohse and Zhang(2015)Lohse, and Zhang]{Detlef2015}
Lohse,~D.; Zhang,~X. {Surface Nanobubbles and Nanodroplets}. \emph{Rev. Mod.
  Phys.} \textbf{2015}, \emph{87}, 981--1035\relax
\mciteBstWouldAddEndPuncttrue
\mciteSetBstMidEndSepPunct{\mcitedefaultmidpunct}
{\mcitedefaultendpunct}{\mcitedefaultseppunct}\relax
\EndOfBibitem
\bibitem[Yu et~al.(2017)Yu, Maheshwari, Zhu, Lohse, and Zhang]{Zhang2017}
Yu,~H.; Maheshwari,~S.; Zhu,~J.; Lohse,~D.; Zhang,~X. {Formation of Surface
  Nanodroplets Facing a Structured Microchannel Wall}. \emph{Lab Chip}
  \textbf{2017}, \emph{17}, 1496--1504\relax
\mciteBstWouldAddEndPuncttrue
\mciteSetBstMidEndSepPunct{\mcitedefaultmidpunct}
{\mcitedefaultendpunct}{\mcitedefaultseppunct}\relax
\EndOfBibitem
\bibitem[Rahimi et~al.(2015)Rahimi, Roberts, Armas-P{\'{e}}rez, Wang,
  Bukusoglu, Abbott, and de~Pablo]{Rahimi2015}
Rahimi,~M.; Roberts,~T.~F.; Armas-P{\'{e}}rez,~J.~C.; Wang,~X.; Bukusoglu,~E.;
  Abbott,~N.~L.; de~Pablo,~J.~J. {Nanoparticle Self-Assembly at the Interface
  of Liquid Crystal Droplets}. \emph{Proc. Natl. Acad. Sci. U. S. A.}
  \textbf{2015}, \emph{112}, 5297--5302\relax
\mciteBstWouldAddEndPuncttrue
\mciteSetBstMidEndSepPunct{\mcitedefaultmidpunct}
{\mcitedefaultendpunct}{\mcitedefaultseppunct}\relax
\EndOfBibitem
\bibitem[Pine(2003)]{Pine2003}
Pine,~D.~J. {Dense Packing and Symmetry in Small Clusters of Microspheres}.
  \emph{Science} \textbf{2003}, \emph{30}, 483--487\relax
\mciteBstWouldAddEndPuncttrue
\mciteSetBstMidEndSepPunct{\mcitedefaultmidpunct}
{\mcitedefaultendpunct}{\mcitedefaultseppunct}\relax
\EndOfBibitem
\bibitem[Arkus et~al.(2009)Arkus, Manoharan, and Brenner]{Arkus2009}
Arkus,~N.; Manoharan,~V.~N.; Brenner,~M.~P. {Minimal Energy Clusters of Hard
  Spheres with Short Range Attractions}. \emph{Phys. Rev. Lett.} \textbf{2009},
  \emph{103}, 118303\relax
\mciteBstWouldAddEndPuncttrue
\mciteSetBstMidEndSepPunct{\mcitedefaultmidpunct}
{\mcitedefaultendpunct}{\mcitedefaultseppunct}\relax
\EndOfBibitem
\bibitem[Comiskey et~al.(1998)Comiskey, Albert, Yoshizawa, and
  Jacobson]{Comiskey1998}
Comiskey,~B.; Albert,~J.~D.; Yoshizawa,~H.; Jacobson,~J. {An Electrophoretic
  Ink for All-Printed Reflective Electronic Displays}. \emph{Nature}
  \textbf{1998}, \emph{394}, 253--255\relax
\mciteBstWouldAddEndPuncttrue
\mciteSetBstMidEndSepPunct{\mcitedefaultmidpunct}
{\mcitedefaultendpunct}{\mcitedefaultseppunct}\relax
\EndOfBibitem
\bibitem[Succi(2001)]{Succi2001}
Succi,~S. \emph{{The Lattice Boltzmann Equation for Fluid Dynamics and
  Beyond}}; Oxford University Press, 2001\relax
\mciteBstWouldAddEndPuncttrue
\mciteSetBstMidEndSepPunct{\mcitedefaultmidpunct}
{\mcitedefaultendpunct}{\mcitedefaultseppunct}\relax
\EndOfBibitem
\bibitem[Shan and Chen(1993)Shan, and Chen]{Shan1993}
Shan,~X.; Chen,~H. {Lattice Boltzmann Model for Simulating Flows with Multiple
  Phases and Components}. \emph{Phys. Rev. E} \textbf{1993}, \emph{47},
  1815\relax
\mciteBstWouldAddEndPuncttrue
\mciteSetBstMidEndSepPunct{\mcitedefaultmidpunct}
{\mcitedefaultendpunct}{\mcitedefaultseppunct}\relax
\EndOfBibitem
\bibitem[Liu et~al.(2016)Liu, Kang, Leonardi, Schmieschek, Narvaez~Salazar,
  Jones, Williams, Valocchi, and Harting]{LKLSNJWVH16}
Liu,~H.; Kang,~Q.; Leonardi,~C.~R.; Schmieschek,~S.; Narvaez~Salazar,~A.;
  Jones,~B.~D.; Williams,~J.~R.; Valocchi,~A.~J.; Harting,~J. {Multiphase
  Lattice Boltzmann Simulations for Porous Media Applications - A Review}.
  \emph{Computat. Geosci.} \textbf{2016}, \emph{20}, 777--805\relax
\mciteBstWouldAddEndPuncttrue
\mciteSetBstMidEndSepPunct{\mcitedefaultmidpunct}
{\mcitedefaultendpunct}{\mcitedefaultseppunct}\relax
\EndOfBibitem
\bibitem[Ladd and Verberg(2001)Ladd, and Verberg]{ladd-verberg2001}
Ladd,~A. J.~C.; Verberg,~R. {Lattice-{B}oltzmann Simulations of Particle-Fluid
  Suspensions}. \emph{J. Stat. Phys.} \textbf{2001}, \emph{104},
  1191--1251\relax
\mciteBstWouldAddEndPuncttrue
\mciteSetBstMidEndSepPunct{\mcitedefaultmidpunct}
{\mcitedefaultendpunct}{\mcitedefaultseppunct}\relax
\EndOfBibitem
\bibitem[Bhatnagar et~al.(1954)Bhatnagar, Gross, and Krook]{Bhatnagar1954}
Bhatnagar,~P.; Gross,~E.; Krook,~M. {A Model for Collision Processes in Gases.
  I. Small Amplitude Processes in Charged and Neutral One-Component Systems}.
  \emph{Phys. Rev.} \textbf{1954}, \emph{94}, 511\relax
\mciteBstWouldAddEndPuncttrue
\mciteSetBstMidEndSepPunct{\mcitedefaultmidpunct}
{\mcitedefaultendpunct}{\mcitedefaultseppunct}\relax
\EndOfBibitem
\bibitem[Chen et~al.(1992)Chen, Chen, and Matthaeus]{Chen1992}
Chen,~H.; Chen,~S.; Matthaeus,~W.~H. {Recovery of the Navier-Stokes Equations
  Using a Lattice-Gas Boltzmann Method}. \emph{Phys. Rev. A} \textbf{1992},
  \emph{45}, R5339\relax
\mciteBstWouldAddEndPuncttrue
\mciteSetBstMidEndSepPunct{\mcitedefaultmidpunct}
{\mcitedefaultendpunct}{\mcitedefaultseppunct}\relax
\EndOfBibitem
\bibitem[Huang et~al.(2007)Huang, Thorne, Schaap, and Sukop]{Huang2007}
Huang,~H.; Thorne,~D.~T.; Schaap,~M.~G.; Sukop,~M.~C. {Proposed Approximation
  for Contact Angles in Shan-and-Chen-Type Multicomponent Multiphase Lattice
  Boltzmann Models}. \emph{Phys. Rev. E} \textbf{2007}, \emph{76}, 066701\relax
\mciteBstWouldAddEndPuncttrue
\mciteSetBstMidEndSepPunct{\mcitedefaultmidpunct}
{\mcitedefaultendpunct}{\mcitedefaultseppunct}\relax
\EndOfBibitem
\bibitem[Aidun et~al.(1998)Aidun, Lu, and Ding]{AIDUN1998}
Aidun,~C.~K.; Lu,~Y.; Ding,~E.-J. {Direct Analysis of Particulate Suspensions
  with Inertia Using the Discrete Boltzmann Equation}. \emph{J. Fluid Mech.}
  \textbf{1998}, \emph{373}, 287--311\relax
\mciteBstWouldAddEndPuncttrue
\mciteSetBstMidEndSepPunct{\mcitedefaultmidpunct}
{\mcitedefaultendpunct}{\mcitedefaultseppunct}\relax
\EndOfBibitem
\bibitem[Stratford et~al.(2005)Stratford, Adhikari, Pagonabarraga, Desplat, and
  Cates]{bib:stratford-adhikari-pagonabarraga-desplat-cates:2005}
Stratford,~K.; Adhikari,~R.; Pagonabarraga,~I.; Desplat,~J.-C.; Cates,~M.~E.
  {Colloidal Jamming at Interfaces: A Route to Fluid-Bicontinuous Gels}.
  \emph{Science} \textbf{2005}, \emph{309}, 2198--2201\relax
\mciteBstWouldAddEndPuncttrue
\mciteSetBstMidEndSepPunct{\mcitedefaultmidpunct}
{\mcitedefaultendpunct}{\mcitedefaultseppunct}\relax
\EndOfBibitem
\bibitem[Joshi and Sun(2009)Joshi, and Sun]{bib:joshi-sun:2009}
Joshi,~A.~S.; Sun,~Y. {Multiphase Lattice {B}oltzmann Method for Particle
  Suspensions}. \emph{Phys. Rev. E} \textbf{2009}, \emph{79}, 066703\relax
\mciteBstWouldAddEndPuncttrue
\mciteSetBstMidEndSepPunct{\mcitedefaultmidpunct}
{\mcitedefaultendpunct}{\mcitedefaultseppunct}\relax
\EndOfBibitem
\bibitem[Ondar\c{c}uhu et~al.(1990)Ondar\c{c}uhu, Fabre, Rapha\"{e}l, and
  Veyssi\'{e}]{Ondurcuhu1990}
Ondar\c{c}uhu,~T.; Fabre,~P.; Rapha\"{e}l,~E.; Veyssi\'{e},~M. {Specific
  Properties of Amphiphilic Particles at Fluid Interfaces}. \emph{J. Phys.}
  \textbf{1990}, \emph{51}, 1527--1536\relax
\mciteBstWouldAddEndPuncttrue
\mciteSetBstMidEndSepPunct{\mcitedefaultmidpunct}
{\mcitedefaultendpunct}{\mcitedefaultseppunct}\relax
\EndOfBibitem
\end{mcitethebibliography}
